\documentclass[aps,onecolumn,floatfix,altaffilletter,preprintnumbers,
               tightenlines,showpacs,showkeys,notitlepage,nofootinbib,superscriptaddress]{revtex4-2}
\usepackage{amsmath}
\usepackage{amssymb}
\usepackage{titlesec}
\usepackage[a4paper,width=165mm,top=25mm,bottom=25mm]{geometry}
\usepackage{slashed}
\usepackage{braket}
\usepackage{bbm}
\usepackage[capitalise]{cleveref}
\usepackage{xcolor}
\usepackage{graphicx}
\usepackage[caption=false,labelformat=simple]{subfig}
\usepackage{siunitx}
\usepackage[force]{feynmp-auto}

\usepackage[normalem]{ulem}

\makeatletter
\newcommand*\rel@kern[1]{\kern#1\dimexpr\macc@kerna}
\newcommand*\widebar[1]{%
  \begingroup
  \def\mathaccent##1##2{%
    \rel@kern{0.8}%
    \overline{\rel@kern{-0.8}\macc@nucleus\rel@kern{0.2}}%
    \rel@kern{-0.2}%
  }%
  \macc@depth\@ne
  \let\math@bgroup\@empty \let\math@egroup\macc@set@skewchar
  \mathsurround\z@ \frozen@everymath{\mathgroup\macc@group\relax}%
  \macc@set@skewchar\relax
  \let\mathaccentV\macc@nested@a
  \macc@nested@a\relax111{#1}%
  \endgroup
}
\makeatother

\begin{document}

\title{Baryogenesis and EDMs in the 2HDM with a Complex Singlet}

\author{Björn Garbrecht}
\email{garbrecht@tum.de}
\affiliation{Physik-Department T70, Technische Universität München, James-Franck-Straße, 85748 Garching, Germany}

\author{Edward Wang}
\email{edward.wang@tum.de}
\affiliation{Physik-Department T70, Technische Universität München, James-Franck-Straße, 85748 Garching, Germany}
\affiliation{Max–Planck–Institut für Physik (Werner–Heisenberg–Institut),
Boltzmannstraße 8, 85748 Garching, Germany}

\preprint{TUM-HEP-1587/25}

\begin{abstract}
\noindent
We perform the first joint analysis of baryogenesis from initial Higgs charges (also called Higgsogenesis) and electric dipole moments (EDMs) in a model with two Higgs doublets, a complex scalar, and a Majorana fermion. In our proposed scenario, baryogenesis happens in three steps: (1) The decay of the scalar produces an asymmetry between the two Higgs doublets which is partly transferred to fermions via Standard Model processes. (2) This asymmetry is converted into a $B-L$ charge via interactions mediated by the Majorana fermion. (3) The weak sphaleron processes partially convert the resulting $B-L$ charge into a $B$ asymmetry. We perform a numerical analysis of baryogenesis and the EDM contributions and find that, due to resonant enhancement, baryogenesis is possible with singlet masses as low as $\SI{e5}{GeV}$. We also find that for some parameters, the model can give large contributions to the electron and neutron EDM while also producing baryogenesis. The present scenario offers a new perspective on interpreting observational bounds in terms of cosmology: Because it relies on out-of-equilibrium decays, the questions of baryogenesis and $CP$ violation in the Higgs sector may be linked with EDM signals without the additional assumption of a first order electroweak phase transition.
\end{abstract}

\maketitle

\section{Introduction}

As first pointed out by Landau~\cite{Landau:1957tp}, permanent electric dipole moments (EDMs) of particles are forbidden in $CP$-conserving theories, making EDMs good probes of $CP$ violation in Nature. While it is now known that $CP$ is not conserved, the only confirmed source of $CP$ violation in the Standard Model (SM) arises from the complex phase in the Cabibbo--Kobayashi--Maskawa (CKM) quark mixing matrix, whose contribution to electron and neutron EDMs are several orders of magnitude below the current experimental sensitivities~\cite{Yamaguchi:2020eub,Dar:2000tn,Seng:2014lea}. Any observation of comparatively large EDMs would therefore be a clear indication of physics beyond the Standard Model.

There has been significant interest in relating baryogenesis to electric dipole moment searches (EDMs), particularly in the context of electroweak baryogenesis, where the same $CP$-violating phases necessary for baryogenesis also give large contributions to the EDMs of the electron and the neutron \cite{Kazarian:1991kb,Aoki:1997tj,Cirigliano:2006dg,Yang:2025mzw}. The first proposals for electroweak baryogenesis were in the contexts of the minimally supersymmetric standard model (MSSM) \cite{Carena:1996wj,Cline:1998hy,Carena:1997gx,Cline:1997vk} and of two-Higgs doublet models (2HDM) \cite{Turok:1990zg,Huet:1995mm,Fromme:2006cm}, but both scenarios are now strongly constrained by LHC data and EDM searches \cite{Cline:2011mm,Cohen:2012zza,Curtin:2012aa}. It was found, however, that adding a scalar singlet can open up the viable parameter space while evading experimental constraints~\cite{Espinosa:2011ax,Espinosa:2011eu,Cline:2012hg}. In addition to this, there have been proposals relating leptogenesis to EDMs~\cite{Chen:2006bv,Joaquim:2007sm,Davidson:2008pf,Kajiyama:2009ae,Abada:2016awd,Borah:2016zbd,Murgui:2025scx}, where the $CP$-violating phases of the Yukawa couplings needed for leptogenesis also give contributions to the EDMs.

The scenario we propose combines features from both electroweak baryogenesis and leptogenesis. We consider a model which extends the Standard Model by an additional Higgs doublet, a complex scalar singlet, and a Majorana fermion. While electroweak baryogenesis was considered for similar models in Refs.~\cite{Davies:1996qn,Alanne:2016wtx,Baum:2020vfl,Biekotter:2025vxl}, we propose here a different mechanism. Baryogenesis in our scenario proceeds in three steps: First, the asymmetric decay of the complex singlet into the Higgs doublets produces a charge which is approximately conserved in the early Universe. In this process, the singlet plays a role in the production of the initial $CP$ asymmetry which is similar to the one of the right-handed neutrino in thermal leptogenesis at the stage of out-of-equilibrium decays. Second, this charge is converted into a $B-L$ charge once the interactions with the Majorana fermion equilibrate. (For illustration of these first two steps, cf. Figure~\ref{fig:time_evolution}.) Third, weak sphaleron processes partly convert this $B-L$ charge into a $B$ charge, as in leptogenesis. The general idea that equilibration mediated by Standard Model singlet fermions could convert an initially $B-L$ symmetric $CP$ asymmetry into a $B-L$ asymmetry has been put forward in Refs.~\cite{Campbell:1992jd,Cline:1993vv,Cline:1993bd,Dick:1999je,Fukugita:2002hu,Garbrecht:2014kda,Fong:2015vna,Garbrecht:2019zaa,Domcke:2020quw}.

The mechanism by which a $CP$ asymmetry is first produced in the Higgs sector and then transferred to the baryons was proposed in Refs.~\cite{Servant:2013uwa,Davidson:2013psa} and dubbed ``Higgsogenesis''. In the scenario considered in Ref.~\cite{Davidson:2013psa}, an asymmetry between the charges of the two Higgs fields is directly converted into a $B$ asymmetry via SM processes, without the need for an intermediate $B-L$ charge. As the authors point out, however, this is only possible in type-I and type-X 2HDMs, but not in type-II or type-Y, because at low temperatures the quark Yukawa couplings would equilibrate the two doublets and therefore wash out any charge asymmetry. In addition to this, there are strong constraints on the flavor-changing scalar couplings of the doublets in order for scalar interactions to be out-of-equilibrium by the time of the electroweak phase transition. In our case, we assume that the transfer of the asymmetry to a $B-L$ charge happens before the up-quark Yukawa interactions, and therefore the Higgs charges, fully equilibrate, so that the mechanism also works for a type-II 2HDM. After this point, the $B-L$ charge is frozen in and insensitive to the remaining interactions with the bath. Because of the relatively high temperatures at which this occurs, the constraints on the doublet potential for our mechanism are considerably weaker than in Ref.~\cite{Davidson:2013psa}.

The outline of this article is as follows: In \cref{sec:model}, we present and discuss some features of the model we consider. In \cref{sec:baryogenesis} we introduce our proposed baryogenesis mechanism, including a discussion of the equilibrated spectator processes at different temperatures, and present the rates of equilibration and for the $CP$-violating source derived in the closed-time-path (CTP) formalism with a careful treatment of the resonant enhancement of the source. In \cref{sec:edms}, we briefly discuss the electron and neutron EDM contributions from our model and show numerical results for randomly sampled parameters in \cref{sec:scan}. We finish with concluding remarks in \cref{sec:conclusion}.

\section{The Model}
\label{sec:model}

In this work, we consider a two-Higgs-doublet model augmented by a complex scalar singlet and a Majorana fermion. The most general scalar sector is of the form 
\begin{align}
	V^d =&\, - \mu_1^2 (\phi_1^\dagger \phi_1) - \mu_2^2 (\phi_2^\dagger \phi_2) - (\mu_{12}^2 (\phi_1^\dagger \phi_2) + h.c.) \notag \\
	& + \lambda_1 (\phi_1^\dagger \phi_1)^2  + \lambda_2 (\phi_2^\dagger \phi_2)^2 + \lambda_3 (\phi_1^\dagger \phi_1) (\phi_2^\dagger \phi_2) + \lambda_4 (\phi_1^\dagger \phi_2) (\phi_2^\dagger \phi_1) \notag \\
	& + (\lambda_5 (\phi_1^\dagger \phi_2)^2 + \lambda_6 (\phi_1^\dagger \phi_1) (\phi_1^\dagger \phi_2) + \lambda_7 (\phi_2^\dagger \phi_2) (\phi_1^\dagger \phi_2) + h.c.), \\
	V^s =& \, - \mu_S^2 (S^* S) - (\mu_S^{\prime 2} (S^2) + h.c.) \notag \\
	& + \lambda_8 (S^* S)^2 + (\lambda_9 (S^* S)(S^2) + \lambda_{10} (S^4) + h.c.) \notag \\
	& + (\kappa_1 (S) + \kappa_2 (S^3) + \kappa_3 (S) (S^* S) + h.c.), \\
	V^{ds} =& \, \lambda_{11} (\phi_1^\dagger \phi_1) (S^* S) + (\lambda_{12} (\phi_1^\dagger \phi_1) (S^2) + \kappa_4 (\phi_1^\dagger \phi_1) (S) + h.c.) \notag \\
	& + \lambda_{13} (\phi_2^\dagger \phi_2) (S^* S) + (\lambda_{14} (\phi_1^\dagger \phi_1) (S^2) + \kappa_5 (\phi_2^\dagger \phi_2) (S) + h.c.) \notag \\
	& + (\lambda_{15} (\phi_1^\dagger \phi_2) (S^* S) + \lambda_{16} (\phi_1^\dagger \phi_2) (S^2) + \lambda_{17} (\phi_1^\dagger \phi_2) (S^{*2}) + h.c.) \notag \\
	& + (\kappa_6 (\phi_1^\dagger \phi_2) (S) + \kappa_7 (\phi_1^\dagger \phi_2) (S^*) + h.c.).
\end{align}
In addition to this, the Yukawa sector is given by
\begin{equation}
	\mathcal{L}_\text{Yukawa} = y_u^i \widebar{Q}_L \widetilde{\phi}_i u_R + y_d^i \widebar{Q}_L \phi_i d_R + y_\ell^i \widebar{L}_L \phi_i e_R + y_\nu^i \widebar{L}_L \widetilde{\phi}_i N + h.c.,
\end{equation}
with $\widetilde{\phi}_i = i \sigma_2 \phi_i^*$ and $N$ a Majorana fermion.

In principle, one could also include additional Majorana fermions, for example in order to explain the neutrino masses. For simplicity, however, we assume that only one of the Majorana fermions equilibrates in the relevant temperature ranges, and do not expect our conclusions to change substantially with the inclusion of additional Majorana fermions provided that they are decoupled from the thermal bath during the production of the $CP$ asymmetry through the decay of the complex singlet. While in a scenario with multiple Majorana fermions, leptogenesis could occur subsequently without involving the singlet, for Majorana masses below the Davidson-Ibarra bound \cite{Davidson:2002qv}, as we consider in the present work, this would in general require quasi-degenerate Majorana fermions, which we do not assume (see Ref.~\cite{Drewes:2022kap} for a recent review).

In general, the couplings of the two Higgs doublets to the Standard Model fermions allow for flavor-changing neutral currents (FCNCs), which are heavily constrained by experiments. One common way of avoiding this is by imposing a $\mathbb{Z}_2$ symmetry for one of the Higgs fields. However, this excludes the possibility of $CP$ violation in the Higgs sector, which might be recovered by a spontaneous or soft breaking of the $\mathbb{Z}_2$ symmetry, e.g. by the $\mu_{12}$ mixing term. Alternatively, one can assume that the Yukawa couplings of the two Higgs doublets are aligned \cite{Pich:2009sp,Alanne:2016wtx}, so that the Yukawa matrices $y^i$ for different Higgs doublets are proportional to each other. While the universal Yukawa alignment might seem arbitrary in the absence of an underlying symmetry, it in fact arises in models of dynamical electroweak symmetry breaking \cite{Simmons:1988fu,Kagan:1991gh,Antola:2009wq,Hashimoto:2009ty,Fukano:2012qx,Geller:2013dla,Alanne:2013dra,Fukano:2013kia,Alanne:2016wtx}.

In our case, we assume that the only source of $CP$ violation in the scalar potential is due to the interactions of the doublets with the singlet. Similar to the type-2 2HDM, we impose a $\mathbb{Z}_2$ symmetry with respect to the transformations
\begin{equation}
	\phi_2 \to - \phi_2, \quad d_R \to - d_R, \quad e_R \to - e_R,
\end{equation}
on the Higgs potential and on the Yukawa sector, so that $\lambda_6 = \lambda_7 = 0$, and up-type particles only couple to $\phi_1$, while down-type ones couple to $\phi_2$. This symmetry is softly broken by the $\mu_{12}$ mass terms and by the couplings of the doublets with the singlet. In our case, however, we additionally assume a large mass hierarchy between the doublet and the singlet sector, so that the $\mathbb{Z}_2$ breaking terms induced by the singlet interactions are suppressed. While one could consider the case $\lambda_5 = \mu_{12} = 0$, so that the $\phi_1$ and $\phi_2$ charges are separately conserved by the doublet potential, this would correspond to an additional global $U(1)$ symmetry, which, when spontaneously broken, would give rise to a massless Goldstone boson.

Of particular importance for the baryogenesis mechanism we propose as well as for the EDM contributions are the cubic singlet-doublet couplings $\kappa_4$ -- $\kappa_7$, while the singlet self-interactions $\lambda_8$ -- $\lambda_{10}$ and the quartic doublet-singlet couplings $\lambda_{11}$ -- $\lambda_{17}$ are not relevant for the leading-order results and are therefore not examined in detail.

After electroweak symmetry breaking (EWSB), $\phi_1$ and $\phi_2$ acquire vacuum expectation values $v_1, v_2$, and it is useful to rotate the doublet fields as
\begin{equation}
	\begin{pmatrix}
		\phi_1 \\
		\phi_2
	\end{pmatrix} = \begin{pmatrix}
		\text{cos} \, \beta & - \text{sin} \, \beta e^{- i \zeta} \\
		\text{sin} \, \beta e^{i \zeta} & \text{cos} \, \beta
	\end{pmatrix}
	\begin{pmatrix}
		H_1 \\
		H_2
	\end{pmatrix},
\end{equation}
with $\text{tan} \, \beta = v_2 / v_1$, so that the vacuum expectation values are entirely contained in $H_1$. In addition to this, the complex singlet $S$ contains a $CP$-even and an odd component, which mix with the neutral components of the two Higgs doublets. In the broken phase, we obtain one charged Higgs $H^+$ as well as two $CP$-even and one $CP$-odd scalar. Following the notation of Refs.~\cite{Haber:2006ue,Boto:2020wyf}, extended to the case with an additional complex singlet, we obtain three $CP$-even ($\varphi^0_{1,2,3}$) and two $CP$-odd ($a^0_{1,2}$) physical neutral states, as well as one charged state ($H^+$):
\begin{equation}
	H_1 = \begin{pmatrix}
		G^+ \\
		\frac{1}{\sqrt{2}} (v + \varphi_1^0 + i G^0)
		\end{pmatrix},
	H_2 = \begin{pmatrix}
		H^+ \\
		\frac{1}{\sqrt{2}} (\varphi_2^0 + i a_1^0)
		\end{pmatrix},
	S = \frac{1}{\sqrt{2}} (\varphi_3^0 + i a_2^0),
\end{equation}
where $G^+$ and $G^0$ are the Goldstone bosons which are ``eaten'' by the $W$ and $Z$ bosons. In this rotated basis, the Lagrangian has the same structure as before with the replacements $\phi \to H, \mu \to Y, \lambda \to Z, \kappa \to K$. The expressions for the parameters in the rotated basis in terms of the original parameters are listed in \cref{sec:params}. We find
\begin{equation}
	m_{H^+}^2 = Y_2 + \frac{1}{2} Z_3 v^2,
\end{equation}
while the squared mass matrix $\mathcal{M}^2$ in the $(\varphi_1^0, \varphi_2^0, a_1^0, \varphi_3^0, a_2^0)$ basis is given by
\begin{align}
	&\mathcal{M}^2 = \nonumber \\
	&\begin{pmatrix}
		v^2 Z_1 & v^2 \text{Re} (Z_6) & - v^2 \text{Im} (Z_6) & \sqrt{2} v \text{Re} (K_4) & -\sqrt{2} v \text{Im} (K_4) \\
		& Y_2 + \frac{v^2}{2} Z_{345}^+ & -\frac{v^2}{2} \text{Im} (Z_5) & \frac{v}{\sqrt{2}} \text{Re} (K_6 + K_7) & - \frac{v}{\sqrt{2}} \text{Im} (K_6 - K_7) \\
		& & Y_2 + \frac{v^2}{2} Z_{345}^- & - \frac{v}{\sqrt{2}} \text{Im} (K_6 + K_7) & - \frac{v}{\sqrt{2}} \text{Re} (K_6 - K_7) \\
		& & & Y_s + \text{Re} (Y_{s1}) + \frac{v^2}{2} (Z_{11} + \text{Re} (Z_{12})) & \frac{1}{2} \text{Im} (2 Y_{s1} + v^2 Z_{12}) \\
		& & & & Y_s - \text{Re} (Y_{s1}) + \frac{v^2}{2} (Z_{11} - \text{Re} (Z_{12}))
	\end{pmatrix},
	\label{eq:mass_matrix}
\end{align}
with $Z_{345}^\pm = Z_3 + Z_4 \pm \text{Re} Z_5$.

This can be diagonalized by an orthogonal matrix $R$ whose elements we parametrize as
\begin{equation}
	R = \begin{pmatrix}
		q_{11} & \text{Re} (q_{12}) & \text{Im} (q_{12}) & \text{Re} (q_{13}) & \text{Im} (q_{13}) \\
		q_{21} & \text{Re} (q_{22}) & \text{Im} (q_{22}) & \text{Re} (q_{23}) & \text{Im} (q_{23}) \\
		q_{31} & \text{Re} (q_{32}) & \text{Im} (q_{32}) & \text{Re} (q_{33}) & \text{Im} (q_{33}) \\
		q_{41} & \text{Re} (q_{42}) & \text{Im} (q_{42}) & \text{Re} (q_{43}) & \text{Im} (q_{43}) \\
		q_{51} & \text{Re} (q_{52}) & \text{Im} (q_{52}) & \text{Re} (q_{53}) & \text{Im} (q_{53})
		\end{pmatrix},
\end{equation}
which obey the orthonormality conditions
\begin{align}
	& \sum_{k=1}^5 q_{k1}^2 = \frac{1}{2} \sum_{k=1}^5 |q_{k2/3}|^2 = 1, \\
	& \sum_{k=1}^5 q_{k2/3}^2 = \sum_{k=1}^5 q_{k2} q_{k3} = \sum_{k=1}^5 q_{k1} q_{k2/3} = 0.
\end{align}

Denoting the mass eigenstates in the broken phase in order of ascending masses by $h_i$, with the respective masses $m_i$, the Yukawa Lagrangian in this new basis is given by
\begin{equation}
\mathcal{L}_\text{Yuk} = - \frac{m_f}{v} \sum_k h_k \bar{f} [q_{k1} - 2 T_3^f c_f \text{Re} (q_{k2}) + i c_f \text{Im} (q_{k2}) \gamma_5] f - \sqrt{2} \left[H^+ \bar{f}' \left(\frac{m_{f'} c_{f'}}{v} P_L + \frac{m_f c_f}{v} P_R \right) V_{f' f} f + c.c. \right],
\end{equation}
where $T_3^f$ is the third component of the weak isospin and $V_{f' f}$ is the CKM matrix. Due to our choice of the $\mathbb{Z}_2$ symmetry transformations, the coupling coefficients correspond to a type-II 2HDM and are given by
\begin{subequations}
\begin{align}
	c_d &= c_\ell = - \text{tan} \, \beta, \\
	c_u &= -\text{cot} \, \beta.
\end{align}
\end{subequations}

Because we assume a large mass hierarchy between the doublet and singlet sectors, we can approximate them as decoupled from one another. In addition to this, because of the $CP$ symmetry of the doublet potential, we can make the approximate identifications $h_1 = h, h_2 = H$, with $h$ the Standard Model Higgs boson and $H$ the additional $CP$-even scalar, as well as $h_3 = A$, the $CP$-odd scalar. After rotating to the mass eigenbasis of the doublets via
\begin{equation}
	\begin{pmatrix}
		h \\
		H
	\end{pmatrix} = \begin{pmatrix}
	\text{cos} \, \theta & - \text{sin} \, \theta \\
	\text{sin} \, \theta & \text{cos} \, \theta
	\end{pmatrix} \begin{pmatrix}
		\varphi_1 \\
		\varphi_2
	\end{pmatrix},
	\label{eq:theta}
\end{equation}
the gauge and Yukawa couplings of the lightest scalar $h$ have a similar form to the SM \cite{Han:2020lta}
\begin{equation}
	\mathcal{L}_{h^0} = \kappa_V \frac{m_Z^2}{v} h Z_\mu Z^\mu + \kappa_V \frac{2 m_W^2}{v} h W_\mu^+ W^{\mu -} - \sum_f \kappa_f \frac{m_f}{v} h \bar{f} f,
\end{equation}
with $f$ denoting the fermionic fields of the Standard Model. Note that, in our notation, the angle $\theta$ corresponds to $- (\beta - \alpha) - \pi/2$ in the standard 2HDM terminology \cite{Arnan:2017lxi,Han:2020zqg}, with the additional field redefinition $\varphi_2 \to - \varphi_2$. In the type-II 2HDM, the coefficients $\kappa$ are given by
\begin{align}
\begin{split}
	\kappa_V &= \text{cos} \, \theta, \\
	\kappa_u &= \text{cos} \, \theta + \text{cot} \, \beta \: \text{sin} \, \theta, \\
	\kappa_{d, \ell} &= \text{cos} \, \theta  + \text{tan} \, \beta \: \text{sin} \, \theta.
\end{split}
\end{align}
In the SM, the coefficients $\kappa$ are all equal to one, which agrees with LHC data \cite{ATLAS:2019nkf,ATLAS:2023vdy,CMS:2026zxu}, and in order to recover this, $\theta$ must be close to $0$ \cite{Atkinson:2021eox}. In this so-called alignment limit $\theta \to 0$, we can express the parameters of the Higgs potential as \cite{Kling:2016opi,Han:2020zqg}
\begin{subequations}
\begin{align}
	v^2 \lambda_1 &= m_h^2 - \frac{\text{tan} \, \beta (\mu_{12}^2 - m_H^2 \, \text{sin} \, \beta \, \text{cos} \, \beta)}{\text{cos}^2 \beta}, \label{eq:lam1}\\
	v^2 \lambda_2 &= m_h^2 - \frac{\mu_{12}^2 - m_H^2 \, \text{sin} \, \beta \, \text{cos} \, \beta}{\text{tan} \, \beta \, \text{sin}^2 \beta}, \label{eq:lam2}\\
	v^2 \lambda_3 &= m_h^2 + 2 m_{H^+}^2 - m_H^2 - \frac{\mu_{12}^2}{\text{sin} \, \beta \, \text{cos} \, \beta}, \label{eq:lam3}\\
	v^2 \lambda_4 &= m_A^2 - 2 m_{H^+}^2 + \frac{\mu_{12}^2}{\text{sin} \, \beta \, \text{cos} \, \beta}, \label{eq:lam4}\\
	v^2 \lambda_5 &= - m_A^2 + \frac{\mu_{12}^2}{\text{sin} \, \beta \, \text{cos} \, \beta}. \label{eq:mA}
\end{align}
\end{subequations}
As mentioned above, setting $\mu_{12} = \lambda_5 = 0$ would give the doublet potential an additional $U(1)$ symmetry, with $A$ being the massless Goldstone boson associated to the breaking of this symmetry, resulting in $m_A^2 = 0$ from \cref{eq:mA}.

Vacuum stability is further assured if \cite{Kling:2016opi}
\begin{equation}
	\mu_{12}^2 = m_H^2 \, \text{sin} \, \beta \, \text{cos} \, \beta, \quad m_h^2 + m_{H^+}^2 - m_H >  0, \quad m_h^2 + m_A^2 - m_H^2 > 0, \label{eq:stability}
\end{equation}
which, while not a strictly necessary condition, we choose to enforce. With this choice, we see from \cref{eq:mA} that $\lambda_5$ is responsible for breaking the mass degeneracy between the $H$ and $A$ scalars. Since, however, current experimental data favors degenerate heavy scalars \cite{Atkinson:2021eox}, we can safely set it to $0$.

\section{Baryogenesis}
\label{sec:baryogenesis}

In this model, we propose a mechanism where baryogenesis occurs in three steps. In a first phase, the out-of-equilibrium decays of the singlet $S$ produces and asymmetry between $\phi_1$ and $\phi_2$, which gets transferred into the different sectors by SM interactions. In a second phase, once the interactions with the Majorana fermion $N$ equilibrate, they convert this asymmetry into a $B-L$ charge. Finally, the $B-L$ charge is partly converted into a $B$ charge by the weak sphaleron processes until the electroweak phase transition happens. After this point, the sphaleron processes are frozen out, and the resulting $B$ charge is conserved.

\subsection{Equilibrium Conditions}

As the early Universe cools, different SM processes equilibrate at different temperatures. While these processes, when only partially equilibrated, can significantly alter the dynamics of leptogenesis \cite{Garbrecht:2014kda,Garbrecht:2019zaa,Garbrecht:2024xfs}, we consider the simplified case where they are either fully decoupled or fully equilibrated. In case partially equilibrated spectators are included, we expect them to protect part of the asymmetry from washout, thus leading to larger final asymmetries. The results of Ref.~\cite{Garbrecht:2014kda} suggest that this effect should be of order one.

For every fully equilibrated process, the sum of the chemical potentials involved vanishes. The relevant SM processes and the equalities enforced are: weak sphaleron
\begin{equation}
	\sum_f (\mu_{L_f} + 3 \mu_{Q_f}) = 0,
\end{equation}
strong sphaleron
\begin{equation}
	\sum_f (2 \mu_{Q_f} - \mu_{u_f} - \mu_{d_f}) = 0,
\end{equation}
lepton Yukawa
\begin{equation}
	- \mu_{e_f} + \mu_{L_f} - \mu_{\phi_2} = 0,
\end{equation}
and quark Yukawa
\begin{align}
	- \mu_{u_f} + \mu_{Q_f} + \mu_{\phi_1} &= 0, \\
	- \mu_{d_f} + \mu_{Q_f} - \mu_{\phi_2} &= 0,
\end{align}
where the $f, f'$ indices for the Yukawa processes label the flavors of the particles that couple to the Higgs doublets via $y^{f f'}$. We can work in a basis where the Yukawa matrices of leptons and up-type quarks $y_u, y_\ell$ are diagonal, while down-type quark Yukawa matrices cannot be simultaneously diagonalized with the up-type ones, giving rise to the Cabbibo-Kobayashi-Maskawa quark mixing matrix.

In addition to this, we can assign a chemical potential to every conserved charge $C$
\begin{equation}
	\mu_C = \sum_i n_i^C g_i \mu_i,
\end{equation}
with the sum $i$ running over the low-energy degrees of freedom. Here, $n_i^C$ are the charge assignments and $g_i$ the multiplicities of the different species. 

We define a basis of particle species
\begin{equation}
	(e, \mu, \tau, L_1, L_2, L_3, u, c, t, d, s, b, Q_1, Q_2, Q_3, \phi_1, \phi_2),
\end{equation}
where the multiplicities are
\begin{equation}
	(g_i) = (1, 1, 1, 2, 2, 2, 3, 3, 3, 3, 3, 3, 6, 6, 6, 4, 4).
\end{equation}
The multiplicities contain isospin and color degrees of freedom, and the factor $2$ from Bose statistics for bosons. The charge asymmetry of each particle species $i$ can be approximated to first order in the chemical potentials as
\begin{equation}
	q_i = g_i \mu_i \frac{T^2}{6},
	\label{eq:charge}
\end{equation}
while for the conserved charges we define
\begin{equation}
	q_C = \mu_C \frac{T^2}{6},
\end{equation}
since the multiplicities are already included in the definition of $\mu_C$. We further define the charge yields as
\begin{equation}
	Y_i = \frac{q_i}{s},
\end{equation}
with $s$ the entropy density.

There are four charges exactly conserved by the Standard Model: weak hypercharge $Y$ and $\Delta_f = B/3 - L_f$. The corresponding charge vectors are
\begin{align}
	n^Y &= (-1, -1, -1, -1/2, -1/2, -1/2, 2/3, 2/3, 2/3, -1/3, -1/3, -1/3, 1/6, 1/6, 1/6, 1/2, 1/2), \\
	n^{\Delta_e} &= (-1, 0, 0, -1, 0, 0, 1/9, 1/9, 1/9, 1/9, 1/9, 1/9, 1/9, 1/9, 1/9, 0, 0), \\
	n^{\Delta_\mu} &= (0, -1, 0, 0, -1, 0, 1/9, 1/9, 1/9, 1/9, 1/9, 1/9, 1/9, 1/9, 1/9, 0, 0), \\
	n^{\Delta_\tau} &= (0, 0, -1, 0, 0, -1, 1/9, 1/9, 1/9, 1/9, 1/9, 1/9, 1/9, 1/9, 1/9, 0, 0).
\end{align}
In addition to this, it is useful to define an additional charge to capture the $CP$ asymmetry produced from the singlet decays. In our scenario, the $CP$ asymmetry is first produced as an asymmetry between $\phi_1$ and $\phi_2$. While the charge difference $\phi_1 - \phi_2$ itself is not conserved by SM processes, we can define the charge
\begin{equation}
	n^{C_\lambda} = (0,0,0,1,1,1,14/3,-4/3,-4/3,-4/3,-4/3,-4/3,-1/3,-1/3,-1/3,-1,1),
\end{equation}
which is conserved except for $y_u^{11}, \lambda_5$ and singlet interactions. The choice of this charge is not unique: adding any linear combination of the other conserved charges would also give a conserved charge.

While up-quark Yukawa interactions only equilibrate at low temperatures, processes mediated by $\lambda_5$ could equilibrate this charge already at higher temperatures. Since, however, as we have discussed in \cref{sec:model}, in our choice of parameters $\lambda_5$ is proportional to the squared mass difference between $A$ and $H$, and current experimental data shows a strong preference for degenerate scalars, we set $\lambda_5$ to $0$ and neglect this effect. As for the singlet interactions, in spite of their suppression by $1/m_S^4$, it is possible that they equilibrate before the Majorana fermion does, if the cubic couplings are large enough. We therefore compute their equilibration rate explicitly and include them as a washout process in our fluid equations in \cref{sec:rates}.

At any given temperature, the conserved charges and equilibrated interactions impose a set of linear equations on the chemical potentials
\begin{equation}
\begin{split}
	\sum_i n_i^I \mu_i &= 0, \\
	\sum_i n_i^C g_i \mu_i &= \mu_C.
\end{split}
\end{equation}
We can solve this set for $\mu_i$ assuming fixed conserved charges $\mu_C$ and denote by the matrices $S$ and $C$ the coefficients of the linear combinations
\begin{equation}
	\mu_{L_f} + \mu_{\phi_1} = \sum_{C \neq \Delta_f} S_{f C} \mu_C - \sum_{f'} C_{f f'} \mu_{\Delta_{f'}},
\end{equation}
which are computed explicitly for different temperature regimes below. When $N$ fully equilibrates, $\mu_{L_f}+\mu_{\phi_1}=0$ because $N$ is a Majorana fermion and it is not possible to assign a conserved chemical potential to it. We can then write
\begin{equation}
	\mu_{\Delta_f} = \sum_{C \neq \Delta_f} \sum_{f'} C_{f f'}^{-1} S_{f' C} \mu_C,
\end{equation}
which describes the transfer of asymmetry from the other conserved charges to $\Delta_f$. After the $N$ interactions freeze-out, these charges are exactly conserved, and contain the $B-L$ asymmetry we observe today.

A few remarks regarding the formalism are in order. First, it is assumed that the right-handed neutrino couples to all three lepton flavors and that its interactions are strong enough to equilibrate all of them. Second, we assume the production of the charge $\mu_{C_\lambda}$ freezes out before the $N$ interactions come into play, so that one mechanism does not interfere with the other. We also only consider fully equilibrated or non-equilibrated processes and neglect partially equilibrated spectators for simplicity, even though partially equilibrated spectators can significantly impact the dynamics of the system \cite{Garbrecht:2014kda,Garbrecht:2019zaa}. Finally, it is necessary to specify the basis of lepton flavors one is working in, depending on the temperature range. At high temperatures, all lepton Yukawa interactions are out of equilibrium, so that flavor effects are irrelevant. We can then work in the basis $(\parallel, \perp_1, \perp_2)$, where $\parallel$ is the vector in flavor space coupling to $N$, while $\perp_1, \perp_2$ are the two directions perpendicular to it. After tau Yukawa interactions come into play, it is necessary to distinguish it from the other two flavors. We then work in the basis $(\parallel_\tau, \tau, \perp)$, where $\ell_{\parallel_\tau} = \ell_\parallel - \frac{\ell_\parallel \cdot \ell_\tau}{\ell_\tau \cdot \ell_\tau} \ell_\tau $ normalized to unity, and $\perp$ is the direction perpendicular to both vectors. Finally, after all lepton Yukawa interactions have equilibrated, we work in the usual $(e, \mu, \tau)$ basis.

The next thing to take into account are the different spectator processes that equilibrate over a range of temperatures. In the present analysis, we assume that the Majorana neutrino $N$ equilibrates before its interactions freeze out at temperatures slightly below its mass. Depending on this freeze-out temperature, following Ref.~\cite{Garbrecht:2014kda}, we identify the following temperature ranges:
\paragraph{$T > \SI{e15}{GeV}$:} all SM interactions are decoupled. We have
\begin{equation}
	\mu_{\Delta_\parallel} = \mu_{B-L} = -\frac{\mu_{C_\lambda}}{8}.
\end{equation}

\paragraph{$\SI{e13}{GeV} < T < \SI{e15}{GeV}$:} only top-quark Yukawa interaction is equilibrated. We find again
\begin{equation}
	\mu_{\Delta_\parallel} = \mu_{B-L} = -\frac{\mu_{C_\lambda}}{8}.
\end{equation}

\paragraph{$10^{11-12} \SI{}{GeV} < T < \SI{e13}{GeV}$:} strong sphalerons are also equilibrated. We find
\begin{equation}
	\mu_{\Delta_\parallel} = \mu_{B-L} = - \frac{7}{60} \mu_{C_\lambda}.
\end{equation}

\paragraph{$\SI{e9}{GeV} < T < 10^{11-12} \SI{}{GeV}$:} weak sphaleron, tau, bottom, and charm Yukawa interactions are equilibrated. In the $(\parallel_\tau, \tau)$ basis, we find
\begin{equation}
	S = \frac{1}{29117} \begin{pmatrix}
	1281 \\
	636
	\end{pmatrix}, \quad C = \frac{1}{29117} \begin{pmatrix}
	15101 & 1224  \\
	1224 & 11859
	\end{pmatrix},
\end{equation}
from which we obtain
\begin{equation}
	\mu_{B-L} = -\frac{257}{2033} \mu_{C_\lambda}.
\end{equation}

\paragraph{$\SI{e6}{GeV} < T < \SI{e9}{GeV}$:} all interactions except for first generation Yukawa interactions are equilibrated. In the $(e, \mu, \tau)$ basis, we have
\begin{equation}
	S = -\frac{1}{1204} \begin{pmatrix}
	51 \\
	36\\
	36 
	\end{pmatrix}, \quad C = \frac{1}{36120} \begin{pmatrix}
	18561 & 1416 & 1416 \\
	1416 & 14456 & 2416 \\
	1416 & 2416 & 14456
	\end{pmatrix},
\end{equation}
from which we obtain
\begin{equation}
	\mu_{B-L} = -\frac{60}{317} \mu_{C_\lambda}.
\end{equation}

Below $\SI{e6}{GeV}$, up-quark Yukawa interactions equilibrate, destroying the $C_\lambda$ charge and the associated asymmetry.

Following the analysis of \cite{Laine:1999wv} for an arbitrary number of Higgs doublets, we find that the conversion coefficient between $B-L$ and $B$ charges during an electroweak crossover is given by
\begin{equation}
	\eta_B = \frac{n_B}{n_\gamma} = 7.04 \, Y_{B-L} \frac{32 + 4 N_\phi}{98 + 13 N_\phi},
\end{equation}
with $N_\phi = 2$. We compare this to the combined CMB and BBN value for the baryon-to-photon ratio from Ref.~\cite{Yeh:2022heq}
\begin{equation}
	\eta_B = 6.040 \pm 0.118 \times 10^{-10}.
\end{equation}

\subsection{Kinetic Equations}

Before EWSB, it is useful to decompose the complex scalar into its $CP$-even and odd components as before
\begin{equation}
	S = \frac{1}{\sqrt{2}} (\varphi_3^0 + i a_2^0).
\end{equation}
We can rotate the fields to the mass basis via
\begin{equation}
	\begin{pmatrix}
		S_1 \\
		S_2
	\end{pmatrix} = \begin{pmatrix}
		\text{cos} \, \theta_S & - \text{sin} \, \theta_S \\
		\text{sin} \, \theta_S & \text{cos} \, \theta_S
	\end{pmatrix} \begin{pmatrix}
		\varphi_3^0 \\
		a_2^0
	\end{pmatrix},
\end{equation}
with
\begin{equation}
	\text{tan} \, \theta_S = - \frac{\text{Im} \mu_S'}{\text{Re} \mu_S'},
\end{equation}
and the masses
\begin{equation}
	m_{S 1/2}^2 = \mu_S^2 \mp 2 |\mu_S'^2|.
\end{equation}

The generation of the initial $CP$ asymmetry can be described by a set of momentum-averaged kinetic equations of the form
\begin{subequations}
\begin{align}
	\frac{d}{d z} Y_{Sa} &= - C_{Sa} (Y_{Sa} - Y_{Sa}^\text{eq}), \\
	\frac{d}{d z} (Y_{\phi_1} - Y_{\phi_2}) &= \sum_a S_a (Y_{Sa} - Y_{Sa}^\text{eq}) - W (Y_{\phi_1} - Y_{\phi_2}),
\end{align}
\end{subequations}
where we use $z = m_{S1}/T$ as a dimensionless time variable and $a$ labels the two real singlet fields. In the above equation, $C_{Sa}$ are the equilibration rates of the scalar singlets, $S_a$ are the asymmetry source terms, and $W$ is the washout term. In addition to these rates, also Standard Model processes change the $\phi_1 - \phi_2$ charge, while only $C_\lambda$ is conserved. We therefore need to relate the conserved charge $C_\lambda$ to $\phi_1 - \phi_2$. This again depends on the other interactions in the thermal bath, leading to a relation
\begin{align}
\mu_{\phi_1} - \mu_{\phi_2} = \kappa \mu_{C_\lambda}.
\end{align}

Using the same framework described above, we find the following regimes:

$T > \SI{e15}{GeV}$:
\begin{equation}
	\mu_{\phi_1} - \mu_{\phi_2} = - \frac{1}{4} \mu_{C_\lambda}.
\end{equation}

$T \in (10^{13}, 10^{15}) \SI{}{GeV}$:
\begin{equation}
	\mu_{\phi_1} - \mu_{\phi_2} = - \frac{5}{24} \mu_{C_\lambda}.
\end{equation}

$T \in (10^{11-12}, 10^{13}) \SI{}{GeV}$: 
\begin{equation}
	\mu_{\phi_1} - \mu_{\phi_2} = - \frac{37}{184} \mu_{C_\lambda}.
\end{equation}

$T \in (10^{9}, 10^{11-12}) \SI{}{GeV}$: 
\begin{equation}
	\mu_{\phi_1} - \mu_{\phi_2} = - \frac{3044}{29117} \mu_{C_\lambda}.
\end{equation}

$T  \in (10^{6}, 10^{9}) \SI{}{GeV}$: 
\begin{equation}
	\mu_{\phi_1} - \mu_{\phi_2} = - \frac{22}{301} \mu_{C_\lambda}.
\end{equation}

We can then rewrite the kinetic equations as
\begin{subequations}
\begin{align}
	\frac{d}{d z} Y_{Sa} &= - C_{Sa} (Y_{Sa} - Y_{Sa}^\text{eq}), \\
	\frac{d}{d z} Y_{C_\lambda} &= \sum_a \frac{1}{\kappa} S_a (Y_{Sa} - Y_{Sa}^\text{eq}) - W  Y_{C_\lambda}. \label{eq:kin_lam}
\end{align}
\end{subequations}

\begin{figure}
\captionsetup[subfloat]{captionskip=4em}
\subfloat[]{
\begin{fmffile}{CP-source_wf1}
\begin{fmfgraph*}(120,50)
\fmfleft{i1,i,i2}
\fmfright{o1,o,o2}
\fmf{scalar, label=$\phi_i$}{i,v1}
\fmf{scalar, label=$\phi_i$, label.side=right}{v2,o}
\fmf{phantom}{i,v1,v2,o}
\fmf{phantom}{i1,vp3,v3,v4,vp4,o1}
\fmffreeze
\fmfshift{(-5,0)}{v3}
\fmfshift{(5,0)}{v4}
\fmf{dbl_dots, left, tension=0.4, label=$S$}{v1,v2}
\fmf{scalar, right=0.3, tension=0.4, label=$\phi_j$}{v1,v3}
\fmf{scalar, right, tension=0.4, label=$\phi_k$}{v3,v4}
\fmf{scalar, right=0.3, tension=0.4, label=$\phi_\ell$}{v4,v2}
\fmf{dbl_dots, left, tension=0.4, label=$S$}{v3,v4}
\end{fmfgraph*}
\end{fmffile}
\label{subfig:wavefunction1}}
\subfloat[]{
\begin{fmffile}{CP-source_wf2}
\begin{fmfgraph*}(120,50)
\fmfleft{i1,i,i2}
\fmfright{o1,o,o2}
\fmf{scalar, label=$\phi_i$}{i,v1}
\fmf{scalar, label=$\phi_i$, label.side=right}{v2,o}
\fmf{phantom}{i,v1,v2,o}
\fmf{phantom}{i2,vp3,v3,v4,vp4,o2}
\fmffreeze
\fmfshift{(-5,0)}{v3}
\fmfshift{(5,0)}{v4}
\fmf{scalar, right, tension=0.4, label=$\phi_j$}{v1,v2}
\fmf{dbl_dots, left=0.3, tension=0.4, label=$S$}{v1,v3}
\fmf{scalar, right, tension=0.4, label=$\phi_k$}{v3,v4}
\fmf{dbl_dots, left=0.3, tension=0.4, label=$S$}{v4,v2}
\fmf{scalar, right, tension=0.4, label=$\phi_\ell$}{v4,v3}
\end{fmfgraph*}
\end{fmffile}
\label{subfig:wavefunction2}}
\caption{Diagrammatic representation of the $CP$-violating source terms.}
\label{fig:CP-source}
\end{figure}

\subsection{Rates}
\label{sec:rates}

The equilibration rates for $S$ and $\phi$ are computed here to leading order in the CTP-formalism, where we include the expansion of the Universe following the procedure from Ref.~\cite{Beneke:2010wd}. Since the relevant couplings for leptogenesis are the cubic couplings between the scalars and the singlet, we define $c_{11} = \tilde{c}_{11}^* = \kappa_4, c_{22} = \tilde{c}_{22}^* = \kappa_5, c_{21} = \kappa_6, \tilde{c}_{21} = \kappa_7$, as well as $c_{12} = \tilde{c}_{21}^*, \tilde{c}_{12} = c_{21}^*$ and call $m_{\phi 1}, m_{\phi 2}$ the effective masses of the two doublets before EWSB. In this notation, $c_{ij}$ mediates the interaction of the complex singlet $S$ with an incoming doublet $\phi_i$ and an outgoing doublet $\phi_j$, while the $\tilde{c}$ couplings mediate the interactions involving $S^*$.

In the singlet mass basis, the cubic couplings of the singlet and the doublets can be written as
\begin{equation}
	c_{ij}^{(a)} (\phi_j^\dagger \phi_i) S_a,
\end{equation}
where we define
\begin{align}
	c_{ij}^{(1)} &= c_{ij} (\text{cos} \, \theta_S - i \, \text{sin} \, \theta_S) + \tilde{c}_{ij} (\text{cos} \, \theta_S + i \, \text{sin} \, \theta_S), \\
	c_{ij}^{(2)} &= c_{ij} (i \, \text{cos} \, \theta_S + \text{sin} \, \theta_S) + \tilde{c}_{ij} (- i \, \text{cos} \, \theta_S + \text{sin} \, \theta_S),
\end{align}
 with the property $c_{ji}^{(a)} = c_{ij}^{(a)*}$.

We then find the tree-level singlet equilibration rate
\begin{equation}
	C_{Sa} = \sum_{j,k} \big|c_{jk}^{(a)} \big|^2 \frac{g_w}{16 \pi} \frac{a_R z}{m_S^3} \frac{K_1 \left(z \right)}{K_2 \left(z \right)},
\end{equation}
where $g_\omega$ is the number of $SU(2)$ degrees of freedom, $a_R = (1/2)m_\text{Pl} (\pi^3 g_\star /45)^{-1/2}$ the comoving temperature (see e.g. Ref.~\cite{Garbrecht:2018mrp}), $m_\text{Pl} = \SI{1.22e19}{GeV}$ is the Planck mass and $g_\star$ is the number of relativistic degrees of freedom, and the washout rate
\begin{equation}
	W_1 = \sum_a \big|c_{12}^{(a)} \big|^2 \frac{1}{16 \pi} \frac{z^3 a_R}{m_S^3} K_1 (z),
\end{equation}
due to inverse decays, where we take the singlets to be quasi-degenerate and approximate their masses by the averaged mass $m_S$. In addition to this leading-order washout rate, there is a washout contribution from two-by-two scattering of the doublets mediated by an off-shell singlet. Even though these only contribute at next-to-leading order in the couplings, because the heavy singlet is off-shell, these contributions are not Boltzmann suppressed at late times when the temperature falls below the singlet mass, so that they can become dominant. We find
\begin{equation}
	W_2 =  \sum_a \frac{g_w}{16 \pi^3} \frac{a_R}{m_S^5} \left[ \big|c_{11}^{(a)} c_{12}^{(a)} \big|^2 + 2 \big|c_{12}^{(a)} c_{12}^{(a)} \big|^2 + \big|c_{12}^{(a)} c_{22}^{(a)}\big|^2 \right],
\end{equation}
where we neglect the momentum dependence of the singlet propagator. Because the couplings are dimensionful, this contribution to the washout rate is approximately scale-invariant, leading to an exponential decay of the asymmetry over a long time range. If the couplings are small enough compared to $m_S$, however, it is possible that the asymmetry gets converted into a $B-L$ charge before getting significantly depleted.

As for the source of the $CP$ asymmetry, we only consider wavefunction-type contributions to it, which exhibit resonant enhancement and are therefore dominant. While the resonant peak of the $CP$ asymmetry due to the mixing of the Higgs doublets is damped by the large gauge-induced propagator widths, the same is not true for the singlet propagator, whose contribution to the $CP$ violating source is strongly enhanced in the case of quasidegenerate singlet masses, which we assume here. There are two wavefunction diagrams that contribute to the source term, shown in \cref{fig:CP-source}. The $CP$ cuts that lead to the asymmetry  here are kinematically forbidden at zero temperature, which is one main of the main reasons why we resort to CTP methods.

In the diagram shown in \cref{subfig:wavefunction1}, $CP$ violation arises from the mixing of the scalar doublets, similar to the scenarios considered in Refs.~\cite{Garbrecht:2012pq,Garbrecht:2012pq,Garbrecht:2024bbo}. We can write its contribution as
\begin{equation}
	S_{\phi i}^{(1)} = c_{ij}^{(a)} c_{ki}^{(a)} \int \frac{d^4 k}{(2 \pi)^4} \frac{d^4 p}{(2 \pi)^4} \frac{d^4 q}{(2 \pi)^4} \delta^4 (k-p-q) (i \Delta_{Sa}^> (q) i \Delta_{\phi i}^< (k) - i \Delta_{Sa}^< (q) i \Delta_{\phi i}^> (k)) i D_{jk} (p),
\end{equation}
where $D_{jk}$ is the one-loop resummed mixed scalar propagator, which we can obtain from \cite{Garbrecht:2011aw,Garny:2011hg,Garbrecht:2012qv}
\begin{multline}
2 k^0 \partial_\eta i D_{\phi 12} + i (m_{\phi 1}^2 - m_{\phi 2}^2) i D_{\phi 12} = - \frac{1}{2} i(\Pi_{\phi 12}^{y >} + \Pi_{\phi 12}^{\lambda >} + \Pi_{\phi 12}^{g >}) i(\Delta_{\phi 11}^< + \Delta_{\phi 22}^<) \\
- \frac{1}{2} \sum_k i (\Pi_{\phi kk}^{y >} + \Pi_{\phi kk}^{\lambda >} + \Pi_{\phi kk}^{g >}) i D_{\phi 12} - < \leftrightarrow >,
\label{eq:kinetic_phi}
\end{multline}
where $\Pi^y, \Pi^\lambda, \Pi^g$ are the one-loop scalar self-energies due to Yukawa, scalar and gauge interactions, respectively, and $< \leftrightarrow >$ stands for the same terms on the right-hand side with the CTP indices reversed. The off-diagonal scalar-mediated self-energy describes the mixing of the two doublets induced by the out-of-equilibrium $S$ fields and is responsible for the $CP$-violating source, while the remaining self-energies contribute to the effective width of the propagator.

Even though we do not specify the masses of the scalar doublets before EWSB, we assume them to be small compared to the mass of the singlet, and their mass splitting to be small compared to their widths. We then use the ansatz \cite{Garbrecht:2011aw}
\begin{equation}
	i D_{\phi 12} (p) = 2 \pi \delta (p^2 - m_\phi^2) \frac{\mu_{\phi 12}}{T} \frac{\text{sign} (p^0) e^{|p^0|/T}}{(e^{|p^0|/T} - 1)^2},
\end{equation}
with the chemical potential $\mu_{\phi 12}$ parametrizing the deviation of $\phi$ from equilibrium, and with the average mass $m_\phi$. We integrate \cref{eq:kinetic_phi} over the momentum and define
\begin{equation}
	n_{\phi 12}^{\pm} = 2 \int_0^{\pm \infty} \frac{d k^0}{2 \pi} \int \frac{d^3 k}{(2 \pi)^3} k^0 i D_{\phi 12} (k),
\end{equation}
to obtain, in the static limit (i.e. dropping the derivative term on the left-hand side of \cref{eq:kinetic_phi}),
\begin{equation}
	\pm i (m_{\phi 1}^2 - m_{\phi 2}^2) n_{\phi 12}^\pm = - B_\phi^{\lambda, \slashed{\text{eq}}} - B_\phi^y n_{\phi 12}^\pm - B_\phi^g (n_{\phi 12}^+ + n_{\phi 12}^-) - B_\phi^{\lambda, \text{even}} n_{\phi 12}^\pm - B_\phi^{\lambda, \text{odd}} n_{\phi 12}^\mp,
	\label{eq:static_phi}
\end{equation}
which has the solution
\begin{subequations}
\begin{align}
	q_{\phi 12} &= n_{\phi 12}^+ - n_{\phi 12}^- = \mathcal{R}_\phi 2 i B_\phi^{\lambda, \slashed{\text{eq}}}, \\
	\mathcal{R}_\phi &= \frac{m_{\phi 1}^2 - m_{\phi 2}^2}{(m_{\phi 1}^2 - m_{\phi 2}^2)^2 + (B_\phi^y + B_\phi^{\lambda, \text{even}} - B_\phi^{\lambda, \text{odd}})(B_\phi^y + 2 B_\phi^g + B_\phi^{\lambda, \text{even}} + B_\phi^{\lambda, \text{odd}})}.
\end{align}
\end{subequations}
The various averaged rates that appear here are defined and estimated in \cref{app:rates}. With this we find the wavefunction contribution to the self-energy
\begin{equation}
	S_{\phi 1}^{(1)} = (c_{11}^{(1)} c_{22}^{(2)} - c_{22}^{(1)} c_{11}^{(2)}) \text{Im} [c_{12}^{(2)} c_{21}^{(1)}] \mathcal{R}_\phi 3 i \frac{z^3 a_R}{128 \pi^4 m_S^3} K_1 (z).
\end{equation}

Similarly, we can write the contribution from the diagram shown in \cref{subfig:wavefunction2} as
\begin{equation}
	S_{\phi i}^{(2)} = c_{ij}^{(a)} c_{ji}^{(b)} \int \frac{d^4 k}{(2 \pi)^4} \frac{d^4 p}{(2 \pi)^4} \frac{d^4 q}{(2 \pi)^4} \delta^4 (k-p-q) (i \Delta_{\phi j}^> (q) i \Delta_{\phi i}^< (k) - i \Delta_{\phi j}^< (q) i \Delta_{\phi i}^> (k)) i D_{S ab} (p),
\end{equation}
where $D_{Sab}$ is the one-loop resummed mixed singlet propagator, which we can obtain from
\begin{equation}
2 k^0 \partial_\eta i D_{S12} + i \Delta m_S^2 i D_{S12} = - \frac{1}{2} i \Pi_{S12}^{\lambda >} i(\Delta_{S1}^< + \Delta_{S2}^<) - \frac{1}{2} (i \Pi_{S1}^{\lambda >} + i \Pi_{S2}^{\lambda >}) i D_{S12} - < \leftrightarrow >,
\label{eq:kinetic_S}
\end{equation}
where $\Pi^\lambda$ is again the self-energy due to scalar interactions and $\Delta m_S^2$ is the difference in squared masses between $S_1$ and $S_2$.

Approximating
\begin{equation}
	i D_{S12} (p) = 2 \pi \delta (p^2 - m_S^2) \frac{\mu_{S12}}{T} \frac{\text{sign} (p^0) e^{|p^0|/T}}{(e^{|p^0|/T} - 1)^2},
\end{equation}
where $\widebar{m}_S$ is the average singlet mass, we can again define
\begin{equation}
	n_{S12}^\pm = 2 \int_0^{\pm \infty} \frac{d k^0}{2 \pi} \int \frac{d^3 k}{(2 \pi)^3} k^0 i D_{S12} (k),
\end{equation}
and integrate \cref{eq:kinetic_S} to find
\begin{equation}
	\pm i \delta m_{S12}^2 n_{S12}^\pm = - B_S^{\lambda, \slashed{\text{eq}}} -  B_S^{\lambda, \text{even}} n_{S12}^\pm - B_S^{\lambda, \text{odd}} n_{S12}^\mp,
	\label{eq:static_S}
\end{equation}
with the solution
\begin{subequations}
\begin{align}
	q_{S12} &= n_{S12}^+ - n_{S12}^- = \mathcal{R}_S 2 i B_S^{\lambda, \slashed{\text{eq}}}, \\
	\mathcal{R}_S &= \frac{\delta m_{S12}^2}{(\delta m_{S12}^2)^2 + (B_S^{\lambda, \text{even}} - B_S^{\lambda, \text{odd}})(B_S^{\lambda, \text{even}} + B_S^{\lambda, \text{odd}})}.
\end{align}
\end{subequations}
Again, the various momentum-averaged rates and their estimates can be found in \cref{app:rates}.

With this we find
\begin{equation}
	S_{\phi 1}^{(2)} = (\text{Im} [c_{12}^{(1)} c_{21}^{(2)}] (c_{11}^{(1)} c_{11}^{(2)} + c_{22}^{(1)} c_{22}^{(2)}) + \text{Im} [c_{12}^{(1),2} c_{21}^{(2),2}]) \mathcal{R}_S \frac{3 a_R z^3}{128 \pi^4 m_S^4} m_S K_1 (z). 
\end{equation}

The two source terms exhibit the same $z$-dependence and can therefore be taken together when computing the final asymmetry. Because of the large widths induced by the gauge interactions of the doublets, we find that $\mathcal{R}_\phi$ is strongly suppressed compared to $\mathcal{R}_S$, so that the second contribution dominates. The kinetic equation for $Y_{C_\lambda}$ \cref{eq:kin_lam} with vanishing initial conditions is formally solved by
\begin{equation}
Y_{C_\lambda} (z) = \sum_a \int_{z_i}^z \frac{2 S_{\phi 1} (z')}{\kappa} (Y_{Sa} - Y_{Sa}^\text{eq}) (z') e^{-\int_{z'}^z W_{\phi} (z'') dz''} d z' = \sum_a \int_{z_i}^z \frac{2 S_{\phi i} (z')}{\kappa \, \mathcal{C}_{Sa} (z')} \frac{d Y_{Sa}}{d z'} e^{-\int_{z'}^z W_{\phi} (z'') dz''} d z',
\end{equation}
which, in the strong washout regime, can be approximated by \cite{Kolb:1983ni}
\begin{equation}
	Y_{C_\lambda} (z) = \sum_a \int_{z_i}^z \frac{2 S_{\phi 1} (z')}{\kappa \, \mathcal{C}_{Sa} (z')} \frac{d Y_{Sa}^\text{eq}}{d z'} e^{-\int_{z'}^z W_{\phi} (z'') dz''} d z'.
\end{equation}

\section{EDMs}
\label{sec:edms}

An important consequence of $CP$ violation in 2HDM models is the prediction of non-vanishing EDMs, particularly of electrons and neutrons. In these models, as first noted by Barr and Zee in Ref.~\cite{Barr:1990vd}, due to the necessary chirality flip, two-loop contributions can be as large or larger than one-loop ones.

The electric dipole moment of a fermion $f$, $d_f$ is given by the coefficient of the effective operator
\begin{equation}
	\mathcal{L}_\text{EDM} = - \frac{d_f}{2} \bar{f} \sigma^{\mu \nu} (i \gamma^5) f F_{\mu \nu},
\end{equation}
with $F_{\mu \nu}$ the electromagnetic field strength tensor.

Since the electron is an elementary particle, its EDM can be readily evaluated using perturbation theory. As for the neutron, since it is a composite particle, its EDM receives contributions from the quark EDMs and chromo EDMs (CEDMs), the Weinberg operator, and four-fermion operators, where the latter two are typically subdominant \cite{Demir:2002gg,Hisano:2012sc}.

Neglecting contributions from the strong $CP$ theta-angle, the neutron EDM in terms of the quark (C)EDMs can be determined using QCD sum rules and is given by \cite{Pospelov:2000bw,Hisano:2012sc}
\begin{equation}
	d_n = 0.47 d_d - 0.12 d_u+ e(-0.18 \widetilde{d}_u + 0.18 \widetilde{d}_d - 0.008 \widetilde{d}_s),
\end{equation}
where $\tilde{d}$ are CEDMs, defined as the coefficient
\begin{equation}
	\mathcal{L}_\text{CEDM} = - \frac{\widetilde{d}_f}{2} g_s \widebar{f} \sigma^{\mu \nu} (i \gamma^5) f G_{\mu \nu},
\end{equation}
with $G_{\mu \nu}$ the gluon field strength tensor and $g_s$ the strong coupling constant.

In Ref.~\cite{Altmannshofer:2020shb}, the full two-loop contribution of 2HDMs with a softly broken $\mathbb{Z}_2$-symmetry to the electron EDM was computed, which was later generalized to unconstrained 2HDMs in Ref.~\cite{Altmannshofer:2025nsl}. The contributions to the EDMs and CEDMs of the light quarks from 2HDMs were computed in Ref.~\cite{Altmannshofer:2025wlc}.

\section{Numerical Results}
\label{sec:scan}

In our parametrization, we first choose the mass eigenstates and the rotation matrix, and from the rotated mass matrix we extract the parameters of the theory. However, the mass matrix contains degeneracies in the parameters, while some parameters do not appear in it at all, and so they need to be fixed separately. In particular, $Y_2 + v^2/2 Z_3, Y_s + v^2/2 Z_{11}$ and $Y_{s1} + v^2/2 Z_{12}$ always appear together, while $Z_2, Z_7, Z_{13} - Z_{17}$ and $K_5$ do not come up at all. In our case, we are only interested in the cubic interactions of the doublets with the singlet and on the doublet scalar interactions, which are given by the $\kappa$ and the $\lambda_i$ parameters. In the limit where the singlets are decoupled from the doublets, we can determine the $\lambda$ parameters from \cref{eq:lam1,eq:lam2,eq:lam3,eq:lam4,eq:mA}. The $\kappa$ parameters can be determined from the $K$ parameters via \cref{eq:K}, but the system is not closed since $K_5$ is not fixed by the mass matrix. To avoid large cancellations, we fix the real and imaginary components of $\kappa_4$ as the geometric mean of the real and imaginary components of $K_4, K_6, K_7$ multiplied by $\pm 10^n$, where the sign $\pm$ is picked randomly and the exponent $n$ is taken from a uniform distribution in $[-1,1]$ for the real and imaginary components separately, so that the system becomes closed.

We further parametrize
\begin{equation}
	\mathcal{M}_\text{diag}^2 = \text{diag} (m_1^2, m_2^2, m_3^2, m_4^2, m_5^2),
\end{equation}
where the masses appear in increasing order, with $m_1 = m_h$ the mass of the physical Higgs boson, and where $m_4$ and $m_5$ are taken to be much heavier than the other three, corresponding to the masses of the singlet-like states. The special orthogonal matrix $R$ that diagonalizes the mass matrix (\ref{eq:mass_matrix})
\begin{align}
	R \mathcal{M}^2 R^T &= \mathcal{M}_\text{diag}^2, \\
	\begin{pmatrix}
		h_1 \\[1pt]
		h_2 \\[1pt]
		h_3 \\[1pt]
		h_4 \\[1pt]
		h_5
	\end{pmatrix} &= R \begin{pmatrix}
		\varphi_1^0 \\[1pt]
		\varphi_2^0 \\[1pt]
		a_1^0 \\[1pt]
		\varphi_3^0 \\[1pt]
		a_2^0
	\end{pmatrix},
\end{align}
can be written in block form as
\begin{equation}
	R = \begin{pmatrix}
		A & B \\
		C & D
		\end{pmatrix},
\end{equation}
where $A$ and $D$ are $3 \times 3$ and $2 \times 2$ square matrices, while $B$ and $C$ are $3 \times 2$ and $2 \times 3$ matrices, respectively. In this notation, $A$ describes the mixing of the doublets among themselves, $D$ of the singlets, and $B$ and $C$ the mixing between the two groups. In order to avoid large cancellations of the contributions of the heavy states on the light masses, the entries of $B$ and $C$ must be at most of order $\sim m_\phi/m_S$. The orthogonality condition on $R$ implies
\begin{equation}
	R^T R = \begin{pmatrix}
		A^T A + C^T C & A^T B + C^T D \\
		B^T A + D^T C & B^T B + D^T D
		\end{pmatrix} = \mathbbm{1}.
\end{equation}
Since the entries in $B$ and $C$ are much smaller than $A$ and $D$, we can neglect their contribution to the diagonal terms, and choose $A \in \text{O} (3)$ and $D \in \text{O} (2)$. Choosing $B$ as a $3 \times 2$ matrix with entries smaller than $m_\phi/m_S$, the requirement that the off-diagonal terms vanish fixes $C = - D B^T A$.

In our case, we assume that the doublet potential is $CP$ conserving, so that the only source of $CP$ violation is the mixing of the doublets with the singlet. This implies $\text{Im} (Z_5) = \text{Im} (Z_6) = \text{Im} (Z_7) = 0$. Since the third row/column of $A$ describes the mixing between the $CP$-even and $CP$-odd Higgs states, we can set these entries to zero, so that $A$ is of the form
\begin{equation}
	A = \begin{pmatrix}
			R (\theta) & 0 \\
			0 & 1
		\end{pmatrix},
\end{equation}
with $R(\theta)$ the $2 \times 2$ rotation matrix with angle $\theta$ described in \cref{eq:theta}.

With this, the rotated mass matrix is
\begin{equation}
	\mathcal{M}^2 = \begin{pmatrix}
		A^T M_\phi^2 A + C^T M_S^2 C & A^T M_\phi^2 B + C^T M_S^2 D \\
		B^T M_\phi^2 A + D^T M_S^2 C & B^T M_\phi^2 B + D^T M_S^2 D
		\end{pmatrix},
		\label{eq:mass_matrix2}
\end{equation}
and the cubic scalar couplings $K$ can be determined by matching the off-diagonal components of the mass matrices in \cref{eq:mass_matrix,eq:mass_matrix2}.

Regarding baryogenesis, the gauge and Yukawa couplings give the doublet propagator a large width, while the width of the singlet propagator is solely determined by scalar interactions, whose contributions are considerably smaller. We find that the $CP$-violating source term from the singlet decays is resonantly enhanced and therefore dominant. Furthermore, we assume that the Majorana fermion interactions equilibrate and freeze out at temperatures between $10^6$ and $\SI{e10}{GeV}$, and that all lepton flavors equilibrate fully. Throughout our analysis, we consider a vanishing initial abundance of singlets.

In the numerical implementation, we pick the best fit values from Ref.~\cite{Atkinson:2021eox} for the doublet masses and mixing angles:
\begin{equation}
	\text{tan} \, \beta = 4.2, \quad m_{H^+} = \SI{2140}{GeV}, \quad m_H = \SI{2180}{GeV}, \quad m_A = \SI{2180}{GeV}, \quad \text{sin} \, \theta = -0.0096,
\end{equation}
with the slight modification that the heavy neutral scalars are taken to be exactly degenerate, so that $\lambda_5 = 0$. In general, however, we do not expect the outcome of baryogenesis to be particularly sensitive to the parameters of the doublet potential, since the leading source of $CP$ violation depends primarily on the cubic doublet-singlet couplings $\kappa_{4,5}$. As for the EDMs, we expect the doublet masses to be inversely correlated to the EDM contributions, as some of the terms are logarithmically enhanced by the scale separation of the doublets and the singlets, see Refs.~\cite{Altmannshofer:2025nsl,Altmannshofer:2025wlc}.

Regarding the singlet sector, we generated 1000 random parameter sets. We picked $\text{log}_{10} (m_S/\unit{GeV}) \in [4, 10]$ and $n \in [1,3]$ assuming uniform distributions. We further set $\Delta m_S^2 = m_S^4/v^2 \times (m_\phi^2/m_S^2)^n$, fixed $D$ as a $2 \times 2$ rotation matrix with random phase in $[0, 2 \pi)$, and picked the entries of the matrix $C$ from a uniform distribution in $[- (m_\phi/m_S)^n, (m_\phi/m_S)^n]$.

\begin{figure}[b]
\centering
\includegraphics[scale=1.]{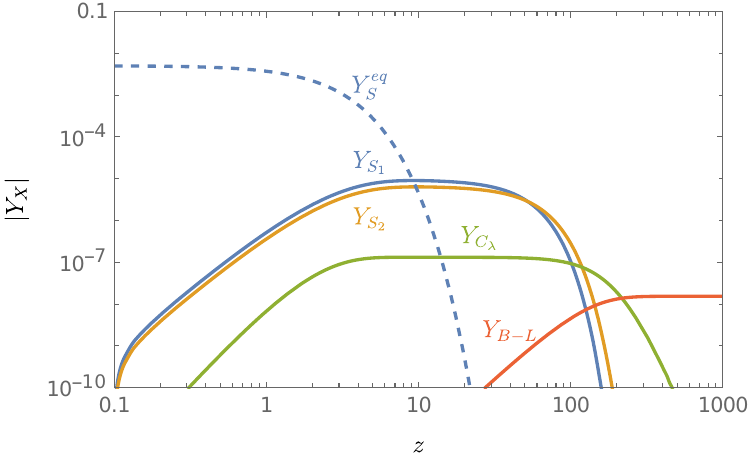}
\caption{Time evolution of the single abundance and of the asymmetries.}
\label{fig:time_evolution}
\end{figure}

For illustrative purposes, we show the time evolution of the singlet abundance and the asymmetries in Figure~\ref{fig:time_evolution} for one typical random parameter point. The asymmetry $Y_{C_\lambda}$ is generated through out-of-equilibrium decays of the singlets. Lepton-number violation mediated by decays and inverse decays of a right-handed neutrino eventually converts this asymmetry into a lepton asymmetry, what leads to a nonvanishing $Y_{B-L}$.

\begin{figure}[t]
\centering
\subfloat[]{
\centering
\includegraphics[scale=0.29]{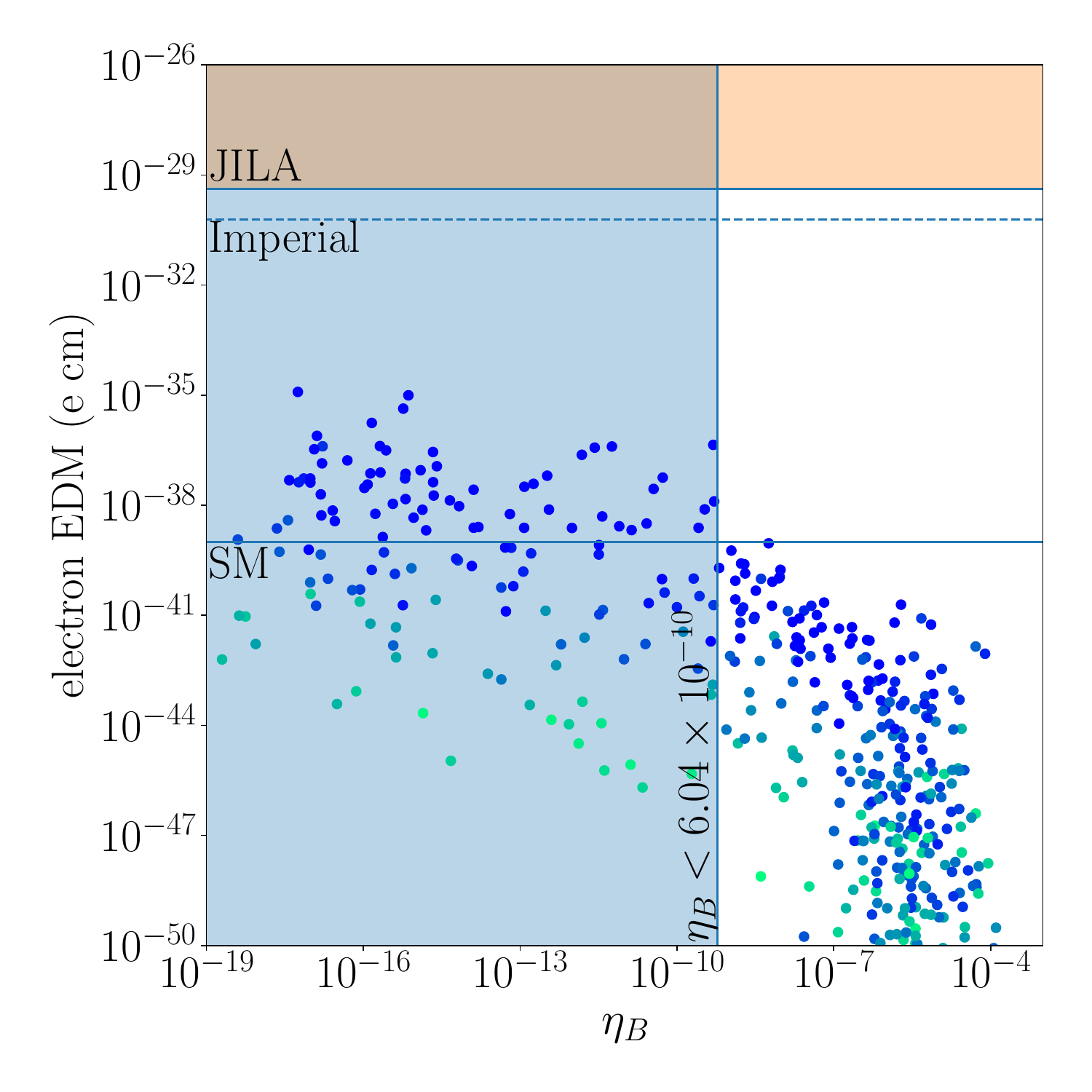}
\label{fig:eEDM}} \hspace{-2em}
\subfloat[]{
\centering
\includegraphics[scale=0.29]{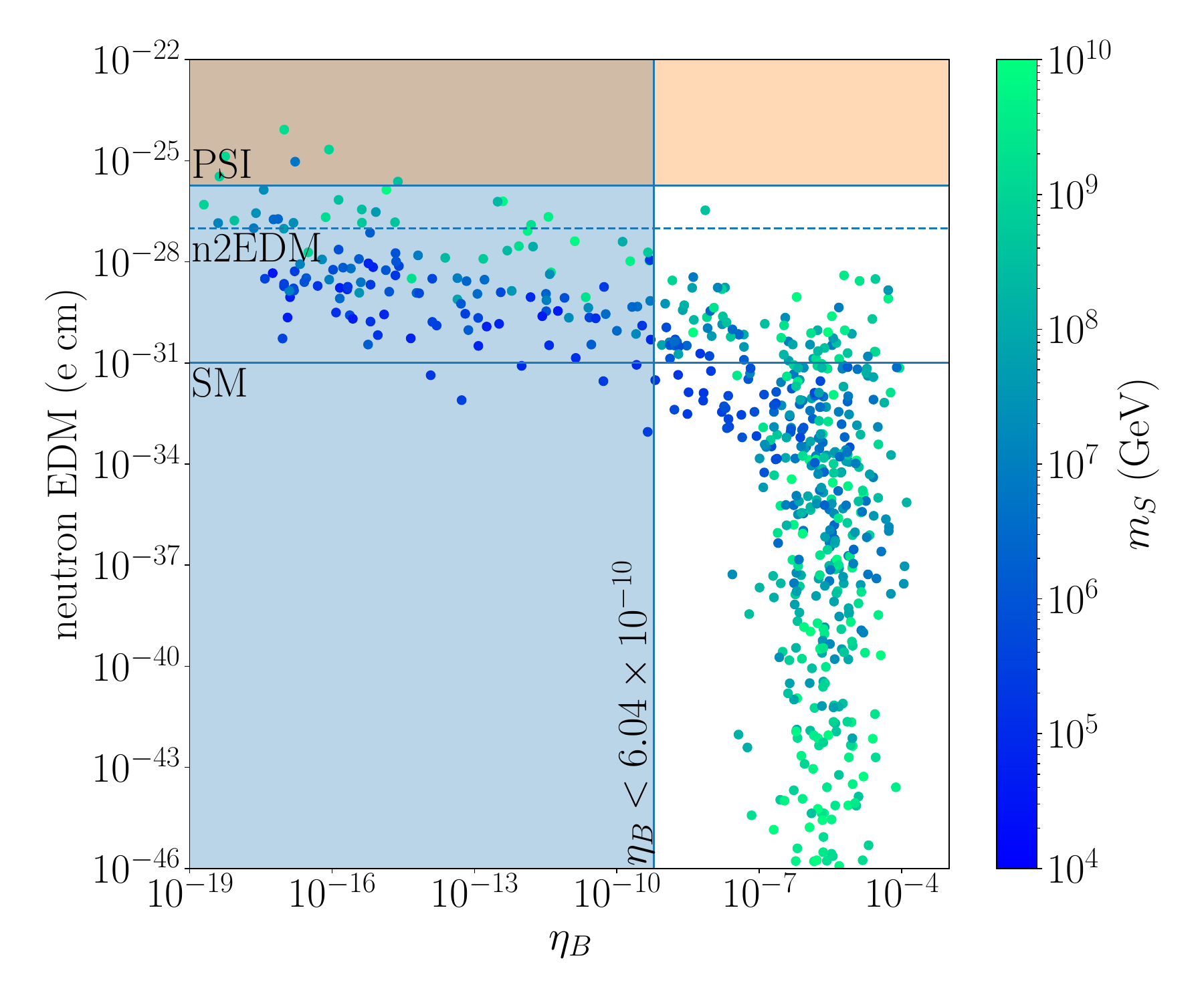}
\label{fig:nEDM}}
\caption{Scan of electron (a) and neutron (b) EDM contributions from our model plotted against the asymmetry parameter $\eta_B$. The blue dashed region is the constraint for successful baryogenesis, while the orange regions are the constraints from JILA \cite{Roussy:2022cmp} for the electron and from PSI \cite{Abel:2020pzs} for the neutron EDMs. Also plotted are the SM predictions for the electron and neutron EDMs from Refs.~\cite{Yamaguchi:2020eub,Dar:2000tn} respectively, and the predicted sensitivities from the experiment under construction at Imperial College \cite{Tarbutt_2013} and the n2EDM experiment at PSI \cite{n2EDM:2021yah}.}
\label{fig:EDM_scan}
\end{figure}

For computing the electron and quark (C)EDMs we made use of a version of the code from Ref.~\cite{Altmannshofer:2025wlc} adapted for the present model. The results of the parameter scan are shown in \cref{fig:EDM_scan}. We compare the electron and neutron EDM contributions from our extended scalar sector with the SM predictions from Refs.~\cite{Yamaguchi:2020eub,Dar:2000tn} respectively, and with the current upper bound from JILA \cite{Roussy:2022cmp} and the predicted sensitivity from Imperial College \cite{Tarbutt_2013} for the electron EDM, as well as the upper bound from PSI \cite{Abel:2020pzs} and the predicted sensitivity of the n2EDM experiment \cite{n2EDM:2021yah} for the neutron EDM.

We find that baryogenesis is possible in this model for singlet masses as low as $\SI{e5}{GeV}$, in large part due to the resonant enhancement of the $CP$-violating source. For such low masses, the singlet decouples before thermalizing, so that the $CP$-asymmetry is produced by the inverse decays of the singlet. We recall that the Majorana fermion must equilibrate at temperatures above $\mathcal{O} (10^6 {\rm GeV})$, since below that the up-quark Yukawa interactions equilibrate the $C_\lambda$ charge. Because we assume that the Majorana fermion equilibrates shortly after the production of the $CP$-asymmetry, the exponential depletion by the 2-to-2 washout processes occurs for only a short time before the $CP$-asymmetry gets converted into a $B-L$ charge, which is protected from washout. We also find that the model can give large contributions to both the electron and the neutron EDMs, dominated by the contributions from the Higgs kinetic terms \cite{Altmannshofer:2025nsl,Altmannshofer:2025wlc}, with some parameter sets already excluded by current neutron EDM constraints \cite{Abel:2020pzs}.

We observe a negative correlation between the EDM contributions and the resulting baryon asymmetry. This is because large EDM contributions require large mixings between the singlets and the $CP$-odd Higgs component, which in turn results from large cubic couplings. Such large cubic couplings, however, also give rise to large washout terms, which eliminates any asymmetry efficiently. Still, we find some parameter sets which can both produce baryogenesis and give sizeable EDM contributions, particularly for the neutron, some of which are just below current experimental bounds. In view of the fast development of EDM experiments, we expect parts of the parameter space of the model to be probed in the near future.

\section{Conclusion}
\label{sec:conclusion}

In this work, we proposed and studied a new mechanism for baryogenesis, where the out-of-equilibrium decays and inverse decays of a thermally produced heavy singlet first produce an asymmetry between the two Higgs doublets, which is converted into a $B-L$ asymmetry via the Majorana fermion interactions and finally converted into a $B$ charge by weak sphaleron processes. Due to its small decay width, the $CP$-violating source is resonantly enhanced, so that baryogenesis is possible for singlet masses as low as $\SI{e5}{GeV}$. We also performed the first combined analysis of baryogenesis and EDMs for a model of baryogenesis from out-of-equilibrium decays with $CP$ violation in the Higgs sector and found that, in spite of the relatively large scale at which $CP$ violation occurs, the additional scalars can give sizeable contributions to the electron and neutron EDMs, which, in the case of the neutron, are just a few orders of magnitude smaller than current experimental sensitivities, and within reach of future experiments. In this way, the mechanism laid out here allows us to directly relate $CP$ violation in the Higgs sector and the resulting EDM contributions to baryogenesis, without the need for a first order electroweak phase transition.

\section*{Acknowledgements}

E.W.'s work is funded by the Deutsche Forschungsgemeinschaft (DFG, German Research Foundation) – SFB 1258 – 283604770.

\appendix

\section{Averaged Rates}
\label{app:rates}

The rates appearing in \cref{eq:static_phi,eq:static_S} are defined as
\begin{align}
	B_\phi^{\lambda, \slashed{\text{eq}}} &= \pm \int_0^{\pm \infty} \frac{d k^0}{2 \pi} \int \frac{d^3 k}{(2 \pi)^3} k^0 i \Pi_{\phi 12}^{\lambda, >} (k) (i \Delta_{\phi 11}^< (k) + i \Delta_{\phi 22}^< (k)) - < \leftrightarrow > \big|_{D_{\phi 12} = 0}, \\
	B_\phi^y &= \pm \frac{1}{n_{\phi 12}^\pm} \int_0^{\pm \infty} \frac{d k^0}{2 \pi} \int \frac{d^3 k}{(2 \pi)^3} k^0 (\Pi_{\phi 11}^{y, \mathcal{A}} (k) + \Pi_{\phi 22}^{y, \mathcal{A}} (k)) i D_{\phi 12} (k), \\
	B_\phi^g &= \pm \frac{1}{n_{\phi 12}^\pm} \int_0^{\pm \infty} \frac{d k^0}{2 \pi} \int \frac{d^3 k}{(2 \pi)^3} 2 k^0 (\Pi_{\phi 11}^{g, \mathcal{A}} (k) + \Pi_{\phi 22}^{g, \mathcal{A}} (k)) i D_{\phi 12} (k), \\
	B_\phi^{\lambda, \text{even}} n_{\phi 12}^\pm + B_\phi^{\lambda, \text{odd}} n_{\phi 12}^\mp &= \pm \int_0^{\pm \infty} \frac{d k^0}{2 \pi} \int \frac{d^3 k}{(2 \pi)^3} k^0 \sum_k (i \Pi_{\phi 12}^{\lambda, >} i \Delta_{\phi kk}^< + i \Pi_{\phi kk}^> i D_{\phi 12} - < \leftrightarrow >)\big|_{\mu_{Sa} = 0}, \\
	B_S^{\lambda, \slashed{\text{eq}}} &= \pm \int_0^{\pm \infty} \frac{d k^0}{2 \pi} \int \frac{d^3 k}{(2 \pi)^3} k^0 i \Pi_{S12}^{\lambda, >} (k) (i \Delta_{S1}^< (k) + i \Delta_{S2}^< (k)) - < \leftrightarrow > \big|_{D_{S 12} = 0}, \\
	B_S^{\lambda, \text{even}} n_{S12}^\pm + B_S^{\lambda, \text{odd}} n_{S12}^\mp &= \pm \int_0^{\pm \infty} \frac{d k^0}{2 \pi} \int \frac{d^3 k}{(2 \pi)^3} \sum_a k^0 (i \Pi_{S12}^{\lambda, >} i \Delta_{Sa}^< + i \Pi_{Sa}^{\lambda, >} i D_{S12} - < \leftrightarrow >)\big|_{\mu_{Sa} = 0},
\end{align}
where we only keep terms up to linear order in the deviations from equilibrium.

We find
\begin{align}
	B_\phi^{\lambda, \slashed{\text{eq}}} &= - \sum_{a,k} c_{1k}^{(a)} c_{k2}^{(a)} \frac{m_S^2 \mu_{Sa}}{32 \pi^3} K_2 (m_S/T), \\
	B_S^{\lambda, \slashed{\text{eq}}} &= g_w \sum_{j,k} c_{jk}^{(1)} c_{kj}^{(2)}\frac{m_S^2 \mu_{Sa}}{32 \pi^3} K_2 (m_S/T).
\end{align}

For $B_\phi^g$ we use the result derived in Ref.~\cite{Garbrecht:2024bbo}
\begin{equation}
	B_\phi^g = 1.4 \times 10^{-3} T^2,
\end{equation}
while for $B_\phi^y$ we only keep the contribution from the coupling of the top quark to $\phi_1$, given by
\begin{equation}
	B_\phi^y = y_t^2 \frac{3}{32 \pi^3} \frac{m_t^4}{T^2} K_2 (m_t/T) \approx y_t^2 m_t^2 \frac{3}{16 \pi^3},
\end{equation}
for $T \gg m_t$.

As for $B_\phi^{\lambda, \text{even/odd}}$, it receives contributions from cubic and quartic couplings, but we only keep the cubic ones, which we expect to be dominant. We then find
\begin{equation}
	B_\phi^{\lambda, \text{even}} = B_\phi^{\lambda, \text{odd}} = \sum_{a, j, k} |c_{jk}^{(a)}|^2 \frac{1}{64 \pi} \frac{m_S^2}{T^2} K_2 (m_S/T).
\end{equation}

Finally, $B_S^{\lambda, \text{even/odd}}$ also receives contributions from cubic and quartic scalar couplings, but we again restrict ourselves to cubic ones. This then gives $B_S^{\lambda, \text{odd}} = 0$ and 
\begin{equation}
	B_S^{\lambda, \text{even}} = \pm \frac{1}{n_{S12}^\pm} \int_0^{\pm \infty} \frac{d k^0}{2 \pi} \int \frac{d^3 k}{(2 \pi)^3} k^0 (\Pi_{S1}^{\lambda, \mathcal{A}} + \Pi_{S2}^{\lambda, \mathcal{A}}) i D_{S12}.
\end{equation}
We find
\begin{equation}
	\Pi_{S1}^{\lambda, \mathcal{A}} + \Pi_{S2}^{\lambda, \mathcal{A}} = g_w \sum_{a,j,k} |c_{jk}^{(a)}|^2 \frac{1}{8 \pi},
\end{equation}
which gives
\begin{equation}
	B_S^{\lambda, \text{even}} = g_w \sum_{a,j,k} |c_{jk}^{(a)}|^2 \frac{1}{16 \pi}.
\end{equation}

\section{Parameters}
\label{sec:params}

Assuming $S$ does not acquire a vacuum expectation value, the minimization condition for the doublet potential gives the relations
\begin{subequations}
\begin{align}
	&\text{Re} (\mu_{12}^2 e^{i \zeta}) v_2 + \mu_1^2 v_1 + \frac{\lambda_1 v_1^3}{2} + \frac{v_2^2 v_1}{2} \lambda_{345} + \frac{3 v_1^2 v_2}{2} \text{Re} (\lambda_6 e^{i \zeta}) + \frac{v_2^3}{2} \text{Re} (\lambda_7 e^{i \zeta}) = 0, \\
	&\text{Re} (\mu_{12}^2 e^{i \zeta}) v_1 + \mu_2^2 v_2 + \frac{\lambda_2 v_2^3}{2} + \frac{v_2 v_1^2}{2} \lambda_{345} + \frac{v_1^3}{2} \text{Re} (\lambda_6 e^{i \zeta}) + \frac{3 v_2^2 v_1}{2} \text{Re} (\lambda_7 e^{i \zeta}) = 0, \\
	& v_1 v_2 \text{Im} (\mu_{12}^2 e^{i \zeta}) + \frac{v_1^2 v_2^2}{2} \text{Im} (\lambda_5 e^{2 i \zeta}) + \frac{v_1^3 v_2}{2} \text{Im} (\lambda_6 e^{i \zeta}) + \frac{v_1 v_2^3}{2} \text{Im} (\lambda_7 e^{i \zeta}) = 0.
\end{align}
\end{subequations}

In addition to this, we can express the couplings in the new basis as
\begin{align}
\begin{split}
	Y_1 &= \mu_1^2 \text{cos}^2 \beta + \mu_2^2 \text{sin}^2 \beta + 2 \text{Re} (\mu_{12}^2 e^{i \zeta}) \text{sin} \beta \text{cos} \beta, \\
	Y_2 &= \mu_2^2 \text{cos}^2 \beta + \mu_1^2 \text{sin}^2 \beta - 2 \text{Re} (\mu_{12}^2 e^{i \zeta}) \text{sin} \beta \text{cos} \beta, \\
	Y_{12} &= e^{- i \zeta} (\text{Re} (\mu_{12}^2 e^{i \zeta}) \text{cos} 2\beta + i \text{Im} (\mu_{12}^2 e^{i \zeta}) + (\mu_2^2 - \mu_1^2) \text{sin} \beta \text{cos} \beta),
\end{split}\\
\begin{split}
	Z_1 &= \lambda_1 \text{cos}^4 \beta + \lambda_2 \text{sin}^4 \beta + 2 \lambda_{345} \text{sin}^2 \beta \text{cos}^2 \beta + 4 \text{Re} (\lambda_6 e^{i \zeta}) \text{sin} \beta \text{cos}^3 \beta + 4 \text{Re} (\lambda_7 e^{i \zeta}) \text{sin}^3 \beta \text{cos} \beta \\
	Z_2 &= \lambda_1 \text{sin}^4 \beta + \lambda_2 \text{cos}^4 \beta + 2 \lambda_{345} \text{sin}^2 \beta \text{cos}^2 \beta - 4 \text{Re} (\lambda_6 e^{i \zeta}) \text{sin}^3 \beta \text{cos} \beta - 4 \text{Re} (\lambda_7 e^{i \zeta}) \text{sin} \beta \text{cos}^3 \beta, \\
	Z_3 &= (\lambda_1 + \lambda_2 - 2 \lambda_{345}) \text{sin}^2 \beta \text{cos}^2 \beta + \lambda_3 - 2 \text{Re} (\lambda_6 e^{i \zeta}) \text{sin} \beta \text{cos} \beta \text{cos} 2 \beta + 2 \text{Re} (\lambda_7 e^{i \zeta}) \text{sin} \beta \text{cos} \beta \text{cos} 2 \beta, \\
	Z_4 &= (\lambda_1 + \lambda_2 - 2 \lambda_{345}) \text{sin}^2 \beta \text{cos}^2 \beta + \lambda_4 - 2 \text{Re} (\lambda_6 e^{i \zeta}) \text{sin} \beta \text{cos} \beta \text{cos} 2 \beta + 2 \text{Re} (\lambda_7 e^{i \zeta}) \text{sin} \beta \text{cos} \beta \text{cos} 2 \beta, \\
	Z_5 &= e^{- 2 i \zeta} [(\lambda_1 + \lambda_2 - \lambda_{345}) \text{sin}^2 \beta \text{cos}^2 \beta + \text{Re} (\lambda_5 e^{i 2\zeta}) + i \text{Im} (\lambda_5 e^{i 2 \zeta}) \text{cos} 2 \beta], \\
	Z_6 &= e^{- i \zeta} [-\lambda_1 \text{sin} \beta \text{cos}^3 \beta + \lambda_2 \text{sin}^3 \beta \text{cos} \beta + \lambda_{345} \text{sin} \beta \text{cos} \beta \text{cos} 2 \beta + i \text{Im} (\lambda_5 e^{2 i \zeta}) \text{sin} \beta \text{cos} \beta \\ 
	&+ \text{Re} (e^{i \zeta} \lambda_6) \text{cos}^2 \beta (2 \text{cos} 2 \beta - 1) + i \text{Im} (e^{i \zeta} \lambda_6) \text{cos}^2 \beta + \text{Re} (e^{i \zeta} \lambda_7) \text{sin}^2 \beta (1 + 2 \text{cos} 2 \beta) + i \text{Im} (e^{- i \zeta} \lambda_7) \text{sin}^2 \beta],\\
	Z_7 &= e^{- i \zeta} [- \lambda_1 \text{sin}^3 \beta \text{cos} \beta + \lambda_2 \text{sin} \beta \text{cos}^3 \beta - \lambda_{345} \text{sin} \beta \text{cos} \beta \text{cos} 2 \beta - i \text{Im} (e^{2 i \zeta} \lambda_5) \text{sin} \beta \text{cos} \beta \\
	&+ \text{Re} (e^{i \zeta} \lambda_6) \text{sin}^2 \beta (1 + 2 \text{cos} 2 \beta) + i \text{Im} (e^{i \zeta} \lambda_6) \text{sin}^2 \beta + \text{Re} (e^{i \zeta} \lambda_7) \text{cos}^2 \beta (2 \text{cos} 2 \beta - 1) + i \text{Im} (e^{i \zeta} \lambda_7) \text{cos}^2 \beta],
\end{split}\\
\begin{split}
	Z_{11} &= \lambda_{11} \text{cos}^2 \beta + \lambda_{13} \text{sin}^2 \beta + \text{Re} (e^{i \zeta} \lambda_{15}) \text{cos} \beta \text{sin} \beta, \\
	Z_{12} &= \frac{1}{2} (\lambda_{12} \text{cos}^2 \beta + \lambda_{14} \text{sin}^2 \beta + e^{i \zeta} \lambda_{16} \text{sin} \beta \text{cos} \beta + e^{- i \zeta} \lambda_{17}^* \text{sin} \beta \text{cos} \beta), \\
	Z_{13} &= \lambda_{11} \text{sin}^2 \beta + \lambda_{13} \text{cos}^2 \beta - \text{Re} (e^{i \zeta} \lambda_{15}) \text{cos} \beta \text{sin} \beta,\\
	Z_{14} &= \frac{1}{2} (\lambda_{12} \text{sin}^2 \beta + \lambda_{14} \text{cos}^2 \beta - e^{i \zeta} \lambda_{16} \text{sin} \beta \text{cos} \beta - e^{- i \zeta} \lambda_{17}^* \text{sin} \beta \text{cos} \beta), \\
	Z_{15} &= e^{- i \zeta} [- \lambda_{11} \text{sin} \beta \text{cos} \beta + \lambda_{13} \text{sin} \beta \text{cos} \beta + \text{Re} (e^{i \zeta} \lambda_{15}) \text{cos} 2 \beta + i \text{Im} (e^{i \zeta} \lambda_{15})], \\
	Z_{16} &= e^{- i \zeta} [- \lambda_{12} \text{sin} \beta \text{cos} \beta + \lambda_{14} \text{sin} \beta \text{cos} \beta + e^{i \zeta} \lambda_{16} \text{cos}^2 \beta - e^{-i \zeta} \lambda_{17}^* \text{sin}^2 \beta],\\
	Z_{17} &= e^{- i \zeta} [- \lambda_{12}^* \text{sin} \beta \text{cos} \beta + \lambda_{14}^* \text{sin} \beta \text{cos} \beta - e^{-i \zeta} \lambda_{16}^* \text{sin}^2 \beta + e^{i \zeta} \lambda_{17} \text{cos}^2 \beta],
\end{split} \\
\begin{split}
	K_{4} &= \kappa_4 \text{cos}^2 \beta + \kappa_5 \text{sin}^2 \beta + e^{i \zeta} \kappa_6 \text{sin} \beta \text{cos} \beta + e^{- i \zeta} \kappa_7^* \text{sin} \beta \text{cos} \beta, \\
	K_{5} &= \kappa_4 \text{sin}^2 \beta + \kappa_5 \text{cos}^2 \beta - e^{i \zeta} \kappa_6 \text{sin} \beta \text{cos} \beta - e^{- i \zeta} \kappa_7^* \text{sin} \beta \text{cos} \beta, \\
	K_{6} &= e^{- i \zeta} [- \kappa_4 \text{sin} \beta \text{cos} \beta + \kappa_5 \text{sin} \beta \text{cos} \beta + e^{i \zeta} \kappa_6 \text{cos}^2 \beta - e^{-i \zeta} \kappa_7^* \text{sin}^2 \beta], \\
	K_{7} &= e^{- i \zeta} [- \kappa_4^* \text{sin} \beta \text{cos} \beta + \kappa_5^* \text{sin} \beta \text{cos} \beta - e^{-i \zeta} \kappa_6^* \text{cos}^2 \beta + e^{i \zeta} \kappa_7 \text{sin}^2 \beta], \label{eq:K}
\end{split}
\end{align}
where $\lambda_{345} = \lambda_3 + \lambda_4 + \text{Re} (\lambda_5 e^{2 i \zeta})$. The phase $\zeta$ can be eliminated by a field redefinition. With the relations above, we find
\begin{equation}
	Y_1 = - \frac{v^2}{2} Z_1, \quad Y_{12} = - \frac{v^2}{2} Z_6.
\end{equation}

\section{EDM Computation}

In this section, we review the methods for computing EDMs at two loops. The EDM can be extracted as the $CP$-odd Pauli form factor of the electromagnetic vertex function
\begin{equation}
	i d_e \bar{u} (p') \sigma^{\mu \nu} \gamma_5 q_\nu u (p),
\end{equation}
in the $q^2 \to 0$ limit. As an example, we consider the contribution from a single Barr-Zee diagram with a top-quark in the loop, shown in \cref{fig:barr-zee}.

\begin{figure}[h]
\begin{fmffile}{Barr-Zee}
\begin{fmfgraph*}(140,90)
\fmfleft{i1,i2,i3,i4}
\fmfright{o1,o2,o3,o4}
\fmf{phantom}{i4,vog,o4}
\fmf{phantom}{i3,vg,o3}
\fmf{phantom}{i2,vh,vg2,o2}
\fmf{phantom}{i1,vhe,vm,vge,o1}
\fmffreeze
\fmfshift{(0,5)}{vg}
\fmfv{l=$\gamma$, l.angle=90}{vog}
\fmf{fermion, label=$e$}{i1,vhe}
\fmf{fermion}{vhe,vge}
\fmf{fermion, label=$e$}{vge,o1}
\fmf{dashes, label=$h$}{vh,vhe}
\fmf{fermion, right=0.56}{vh,vg2,vg,vh}
\fmf{boson, label=$\gamma$, label.side=right}{vge,vg2}
\fmf{boson}{vg,vog}
\end{fmfgraph*}
\end{fmffile}
\caption{Barr-Zee diagram contribution to the electron EDM.}
\label{fig:barr-zee}
\end{figure}

Overall we have four diagrams of the same type, which can be obtained by switching the arrow of the fermion flow in the loop or by exchanging the internal Higgs and photon lines. The four diagrams give the same contribution to the EDM, and so we get a factor of four in the final result. The contribution of a single diagram to the vertex function is given by
\begin{align}
	&V^\mu (p_1, p_2) = i (-i e)^3 N_c (Q_e^t)^2 \frac{(-i m_e) (-i m_t)}{v^2} \nonumber \\
	&\times \int \frac{d^4 k}{(2 \pi)^4} \frac{d^4 k'}{(2 \pi)^4} (q_{k1} - 2 T_3^\ell c_\ell \text{Re} (q_{k2}) + i c_\ell \text{Im} (q_{k2}) \gamma^5) \frac{i (\slashed{k} + m_e)}{k^2 - m_e^2} \gamma^\nu  \frac{-i g_{\rho \nu}}{(p_1-k)^2} \frac{i}{(p_2-k)^2 - m_k^2} \nonumber \\
	&\times\text{tr} \left[ \gamma^\rho \frac{i (\slashed{k}' + m_t)}{k'^2 - m_t^2} \gamma^\mu \frac{i (\slashed{k}' - \slashed{p}_2 + \slashed{p}_1 + m_t)}{(k'-p_2+p_1)^2 - m_t^2} (q_{k1} - 2 T_3^f c_f \text{Re} (q_{k2}) + i c_f \text{Im} (q_{k2}) \gamma^5) \frac{i (\slashed{k}' + \slashed{p}_1 - \slashed{k} + m_t)}{(k'+p_1-k)^2 - m_t^2} \right],
\end{align}
where $Q_e^t$ is the electric charge of the top quark, and $N_c$ is the number of color degrees of freedom.

The first step in evaluating this term is to compute the top-loop vertex function. Gauge invariance via Ward identities implies that this vertex function must be a linear combination of the tensors
\begin{align}
	P^{\rho \mu} &= \epsilon^{\rho \mu \alpha \beta} q_{\alpha} \ell_\beta, \\
	T^{\rho \mu} &= q^\rho \ell^\mu - g^{\rho \mu} q \cdot \ell,
\end{align}
where $\ell = p_1 - k$ is the momentum of the internal photon and $q = p_2 - p_1$ from the external one. We find that the loop function can be decomposed into one part
\begin{align}
	&4 e^2 \frac{m_t^2}{v} c_f \text{Im} (q_{k2}) P^{\rho \mu} \int \frac{d^4 k'}{(2 \pi)^4} \frac{1}{(k'^2 - m_t^2) ((k' + \ell)^2 - m_t^2) ((k'-q)^2 - m_t^2)} \nonumber \\
	\equiv & 4 e^2 \frac{m_t^2}{v} c_f \text{Im} (q_{k2}) P^{\rho \mu} I_1 (\ell^2, q^2),
\end{align}
and one part
\begin{align}
	& 4 e^2 \frac{m_t^2}{v} (q_{k1} - 2 T_3^f c_f \text{Re} (q_{k2})) \int \frac{d^4 k'}{(2 \pi)^4} \frac{-g^{\rho \mu} (k'^2 + l \cdot q - m_t^2) + 2 k'^\mu \ell^\rho + 4 k'^\mu k'^\rho - 2 k'^\rho q^\mu - \ell^\rho q^\mu + \ell^\mu q^\rho}{(k'^2 - m_t^2) ((k' + \ell)^2 - m_t^2) ((k'-q)^2 - m_t^2)} \nonumber \\
	\equiv & \, 4 e^2 \frac{m_t^2}{v} (q_{k1} - 2 T_3^f c_f \text{Re} (q_{k2})) T^{\rho \mu} I_2 (\ell^2, q^2).
\end{align}
Using Feynman parameters and taking the $q^2 \to 0$ limit, we find in dim. reg.
\begin{align}
	I_1 (\ell^2, 0) &=- i (4 \pi)^{\epsilon - 2} \Gamma (\epsilon + 1) \int_0^1 dt \frac{(1-t)}{((t-1) t \ell^2 + m_t^2)^{1 + \epsilon}} \nonumber \\
	&= \frac{i}{16 \pi^2} \int_0^1 dt \frac{1}{t} \frac{1}{(\ell^2 - \tilde{m}_t^2)}, \\
	I_2 (\ell^2, 0) &= i \epsilon (4 \pi)^{\epsilon - 2} \Gamma (\epsilon) \int_0^1 dt \frac{(2 t^3 - 4 t^2 + 3 t - 1) [(t-1) t]^{-1-\epsilon} }{(\ell^2 - \tilde{m}_t^2)^{1 + \epsilon}} \nonumber \\
	&= \frac{i}{16 \pi^2} \int_0^1 dt \frac{2 t^2 - 2 t + 1}{t} \frac{1}{(\ell^2 - \tilde{m}_t^2)},
\end{align}
where
\begin{equation}
	\tilde{m}_t^2 = \frac{m_t^2}{(1-t) t}.
\end{equation}
While one could perform the $t$ integral right away, it is more convenient to leave it unevaluated and treat the denominator as a propagator \cite{Eeg:1983mt}. Making the substitution $\ell \to p_1 - k$ and taking the limit $m_e, p_1, p_2 \to 0$, we find
\begin{align}
	V^\mu (0, 0) =& \, -4 e^3 N_c (Q_e^t)^2 \frac{m_e m_t^2}{v^2} \int \frac{d^4 k}{(2 \pi)^4} (q_{k1} - 2 T_3^\ell c_\ell \text{Re} (q_{k2}) + i c_\ell \text{Im} (q_{k2}) \gamma^5) \frac{\slashed{k}}{k^2} \gamma_\rho \frac{1}{k^2} \frac{1}{k^2 - m_k^2} \nonumber \\
	&\times [(q_{k1} - 2 T_3^f c_f \text{Re} (q_{k2})) (q^\rho k^\mu - g^{\rho \mu} q \cdot k) I_2 ((p_1 - k)^2, 0) + c_f \text{Im} (q_{k2}) \epsilon^{\rho \mu \alpha \beta} q_\alpha k_\beta I_1 ((p_1 - k)^2, 0)].
\end{align}
With
\begin{equation}
	\int \frac{d^4 k}{(2 \pi)^4} \frac{k^\mu k^\nu}{(k^2)^2 (k^2 - m_k^2) (k^2 - \tilde{m}_t^2)} = - \frac{i g^{\mu \nu}}{64 \pi^2 (m_k^2 - \tilde{m}_t^2)} \text{log} \left(\frac{m_k^2}{\tilde{m}_t^2} \right),
\end{equation}
multiplying by four, and keeping only the mixed terms, which give contributions to the EDM, we find
\begin{align}
	d_e =& \, -\frac{32 e^3 m_e m_t^2 N_c (Q_e^t)^2}{1024 \pi^4 v^2} \Bigg[c_\ell \text{Im} (q_{k2}) (q_{k1} - 2 T_3^f c_f \text{Re} (q_{k2})) \int_0^1 dt \frac{2 t^2 - 2 t + 1}{t} \frac{1}{m_k^2 - \tilde{m}_t^2} \text{log} \left(\frac{m_k^2}{\tilde{m}_t^2} \right) \nonumber \\
	&+ (q_{k1} - 2 T_3^\ell c_\ell \text{Re} (q_{k2})) c_f \text{Im} (q_{k2}) \int_0^1 dt \frac{1}{t} \frac{1}{m_k^2 - \tilde{m}_t^2} \text{log} \left(\frac{m_k^2}{\tilde{m}_t^2} \right) \Bigg].
\end{align}
Defining the Davydychev-Tausk vacuum integral function \cite{Davydychev:1992mt}
\begin{align}
	\nonumber \Phi (x,y) =& \, \text{Re} \left\{ \frac{2}{\sqrt{\lambda (x,y)}} \left[ \frac{\pi^2}{6} - \frac{1}{2} \text{log} (x) \text{log} (y) + \text{log} \left(\frac{1+x-y-\sqrt{\lambda(x,y)}}{2} \right) \text{log} \left(\frac{1-x+y-\sqrt{\lambda (x,y)}}{2} \right) \right. \right. \\
	& \left. \left. - \text{Li}_2 \left(\frac{1+x-y-\sqrt{\lambda}}{2} \right) - \text{Li}_2 \left( \frac{1-x+y-\sqrt{\lambda (x,y)}}{2} \right) \right] \right\}
\end{align}
where $\text{Li}_2$ is the dilogarithm and $\lambda (x,y) = (1-x-y)^2 - 4 x y$ is the Källén polynomial, at equal arguments we find
\begin{equation}
	\Phi (x) = \lim_{y \to x} \Phi (x,y) = \frac{2}{\sqrt{1-4 x}} \left[ \frac{\pi^2}{6} + \text{log}^2 \left(\frac{1-\sqrt{1-4x}}{2} \right) - \frac{\text{log}^2 (x)}{2} - 2 \text{Li}_2 \left( \frac{1-\sqrt{1-4x}}{2} \right) \right].
\end{equation}

With this, we can compute the remaining $t$ integrals to find
\begin{align}
	d_e =& - \frac{m_e e^3 m_t^2 N_c (Q_e^t)^2}{64 \pi^4 v^2} c_\ell \sum_k \text{Im} (q_{k2}) \Bigg\{ c_f (q_{k1} - 2 T_3^\ell c_\ell \text{Re} (q_{k2})) r_k \Phi (r_k) \nonumber \\
	& + (q_{k1} - 2 T_3^f c_f \text{Re} (q_{k2})) c_\ell r_k (4 + 2 \text{log} (r_k) + (1 - 2 r_k) \Phi (r_k)) \Bigg\},
\end{align}
in agreement with Refs. \cite{Barr:1990vd,Altmannshofer:2020shb,Altmannshofer:2025nsl}.

\bibliography{references}

\end{document}